\documentclass[aps, prd, 10pt, twocolumn, superscriptaddress,noshowpacs, preprintnumbers, longbibliography,nofootinbib,bibnotes,floatfix]{revtex4-2}
\pdfoutput=1 

\usepackage{amsmath}
\usepackage{amsfonts}
\usepackage{amssymb}
\usepackage{mathtools}
\usepackage{graphics}
\usepackage{tabularx}
\usepackage[export]{adjustbox}
\usepackage{multirow}
\usepackage{citesort}
\usepackage{graphicx}
\usepackage{subfig}
\usepackage{url}
\usepackage{soul}

\usepackage{bm}
\usepackage[dvipsnames]{xcolor}
\usepackage[utf8]{inputenc}

\usepackage[normalem]{ulem}
\usepackage[colorlinks=true,breaklinks=true]{hyperref}
\hypersetup{allcolors=[rgb]{0.0 0.0 1.0},linkcolor=[rgb]{0.75 0.05 0.05}}

\usepackage{tikz,xcolor,hyperref}

\definecolor{lime}{HTML}{A6CE39}
\DeclareRobustCommand{\orcidicon}{\hspace{-1mm}
	\begin{tikzpicture}
	\draw[lime, fill=lime] (0,0) 
	circle [radius=0.16] 
	node[white] {{\fontfamily{qag}\selectfont \tiny \,ID}};
	\draw[white, fill=white] (-0.0525,0.095) 
	circle [radius=0.007];
	\end{tikzpicture}
	\hspace{-3mm}
}

\foreach \x in {A, ..., Z}{\expandafter\xdef\csname orcid\x\endcsname{\noexpand\href{https://orcid.org/\csname orcidauthor\x\endcsname}
			{\noexpand\orcidicon}}
}

\begin{document}

\title{The Interplay of Symmetry Energy Uncertainties, Nonlinear $\sigma-\delta$ Coupling, and Dark Matter in Neutron Star Macroscopic Properties}
\author{Zhaohui Feng\orcidA{}}
\email{fengzh@mails.ccnu.edu.cn}

\author{Xiaoxuan Zhai\orcidC{}}

\author{Shuangxuan Chen\orcidD{}}

\author{Defu Hou\orcidB{}}
\email{houdf@mail.ccnu.edu.cn}

\affiliation{Institute of Particle Physics and Key Laboratory of Quark and Lepton Physics(MOE), Central China Normal University, Wuhan 430079, China}
\begin{abstract}
The poorly constrained density dependence of the nuclear symmetry energy introduces significant uncertainties in the equation of state (EOS) of dense nuclear matter and, consequently, in neutron-star properties. We systematically investigate how uncertainties in the symmetry energy $E_{\rm sym}$ and its slope $L$ at saturation density $n_0$ affect NS properties within the relativistic mean-field (RMF) framework, including the effects of nonlinear $\sigma$-$\delta$ coupling and possible admixture of dark matter(DM) . For fixed $E_{\rm sym}(n_0)$ and $L(n_0)$, we find that the $\sigma$-$\delta$ coupling with $g_{\sigma\delta}=-0.004$ induces an abnormal softening of the EOS, which simultaneously increasing the maximum NS mass and reducing the stellar radius and tidal deformability. A similar behavior is found in the presence of DM with Fermi momentum $k_F^{\rm DM}=50~\mathrm{MeV}$ at $E_{\rm sym}(n_0)=36~\mathrm{MeV}$. In this case, the RMF EOS satisfies the tidal-deformability constraint from GW170817. We further find that the surface curvature of NSs is strongly correlated with the stiffness of $E_{\rm sym}$, with softer symmetry energy corresponding to larger surface curvature. Our results also indicate that the behavior of $E_{\rm sym}$ around $n_0$ is mainly governed by isoscalar rather than isovector parameters.
\end{abstract}
\maketitle

\section{Introduction}

Neutron stars(NS) are remnants of gravitational collapse at the end of stellar evolution. As the densest observable objects in the Universe, NS typically possess masses of 1$\sim$3 solar masses($M_{\odot}$) and radii of approximately 10~km. The properties of dense nuclear matter(NM) in the interiors of the NS have become a major focus of research in nuclear physics, particle physics and astrophysics. A common approach to probe the properties of NM through NS is to solve the Tolman-Oppenheimer-Volkoff(TOV) equations\cite{tolman1939static,oppenheimer1939massive} and obtain the corresponding mass-radius(M-R) relations. The key input in such calculations is the equation of state(EOS) of dense nuclear matter. A realistic EOS must also satisfy the constraints derived from empirical data and terrestrial nuclear experiments(exp/emp)\cite{gambhir1990relativistic, ring1996relativistic, dutra2014relativistic}. As a natural laboratory, NS provide unique environments for studying extreme physics. In particular, the binary neutron star event GW170817\cite{170817} has introduced a new dimension for constraining the EOS of dense-matter through multimessenger observations.

RMF theory has been widely employed to describe finite nuclei and the bulk properties of NM\cite{johnson1955classical, durr1956relativistic, walecka1974theory}. Constrained by both terrestrial nuclear experiments and astrophysical observations, RMF allows for extrapolation of the NM parameters to higher or lower densities essential in the analysis of NS physics. A crucial quantity among the NM properties is the symmetry energy($E_{\text{sym}}\left(n_B\right)$) at the saturation density($n_0$) of symmetric nuclear matter(SNM). Currently, $n_0$ is widely accepted to be around $ 0.16 ~\mathrm{fm}^{-3}$. The symmetry energy and its slope($L(n_B)$) have been extensively investigated through both theoretical calculations and nuclear experiments\cite{li2013constraining}\cite{wang2025extended}, $(E_{\text{sym}})_{n_0}$ is generally found to be around 30~MeV, whereas the corresponding symmetry energy slope~($L_{n_0}$) spans a larger range from several tens to more than one hundred MeV~\cite{li2013constraining}. Studies have shown that $L_{n_0}$ is positively linear correlated with the neutron-skin thickness $(\Delta r_{\mathrm{np}} \equiv r_n-r_p)$ of finite nuclei\cite{brown2000neutron, chen2005nuclear, centelles2009nuclear, typel2001neutron},
where $r_n(r_p)$ is the root-mean-square radii(rms) of the neutron(proton) distribution in the nucleus. The updated Lead Radius EXperiment (PREX-II)\cite{Pb208} reported that neutron skin thickness of ${}^{208} \mathrm{Pb}$, $\Delta r_{\mathrm{np}}=0.283 \pm 0.071 \mathrm{fm}$, generally implies a relatively large $L_{n_0}$. This result is in direct tension with the CREX for ${ }^{48} \mathrm{Ca}$\cite{adhikari2022precision}, $\Delta r_{n p}=0.121 \pm 0.026(\exp ) \pm 0.024 \text { (model) fm }$, which instead favors a smaller value of $L_{n_0}$.

Within the RMF framework, Li et al.\cite{Li:2022okx} used Bayesian inference based on the accurately calibrated FSUGold interaction\cite{FSUGold} to obtain two sets of RMF model parameters including the isoscalar-scalar meson and isovector-scalar meson coupling($\sigma$-$\delta$ coupling)\cite{zabari2019influence}. These parameter sets soften the symmetry energy in the intermediate-densities while stiffening it at high densities. As a consequence, the tension between the tidal deformability constraint of the canonical NS(1.4$M_{\odot}$) $642 \lesssim \Lambda_{1.4} \lesssim 955$, inferred from PREX-II neutron-skin thickness measurements within RMF\cite{reed2021implications} and the upper bound $\Lambda_{1.4} \lesssim 580$ required by GW170817\cite{170817} was alleviated. The $\delta$ meson possesses the SU(2) isospin symmetry within the RMF theory. As an isovector-scalar meson, its coupling to nucleons enters the nucleon effective mass in the Dirac equation, leading to a proton-neutron effective mass splitting in asymmetric NM, particularly at high densities. This feature makes the $\delta$ meson especially suitable for describing neutron-rich matter inside the NS. In the original RMF theory developed by Walecka, Teller, and Duerr\cite{johnson1955classical, durr1956relativistic, walecka1974theory}, 
the Lagrangian included only nucleons together with the $\sigma$ meson and $\omega$(isoscalar-vector) meson. Boguta and Bodmer introduced cubic and quartic self-interaction of $\sigma$ to better control the incompressibility($K$) and the nucleon effective mass($m^*$) at $n_0$\cite{boguta1977relativistic}.

The $\delta$ meson was first introduced into the RMF theory by Kubis et al.\cite{kubis1997nuclear}, Zabari et al.\cite{zabari2019influence} demonstrated that the $\sigma$-$\delta$ coupling can simultaneously soften $E_{\text{sym}}$ in intermediate-density while stiffening it at high densities. Later, the $\sigma$-$\delta$ interaction was further studied by Li et al.\cite{Li:2022okx}, which alleviated the tension between GW170817 and PREX-II as disscussed earlier. Moreover, the RMF parameter set NL3\cite{lalazissis1997new} which itself fails to satisfy the constraints from GW170817, can become compatible with the corresponding EOS constraints once the possible admixture of dark matter(DM) is taken into account\cite{Das:2018frc}.

In the Universe, nearly 85$\%$ of the matter is believed to consist of DM, which has not yet been directly detected. Using observations from the James Webb Space Telescope (JWST), Scognamiglio et al.\cite{scognamiglio2026ultra} recently presented one of the highest-resolution maps of the DM distribution to date. In addition, the existence of DM has been inferred from a wide range of astrophysical and cosmological observations, including galaxy clusters, galactic rotation curves, cosmic microwave background anisotropies, and many others. Some studies suggest that DM particles may possess weak self-interactions and may be captured by the strong gravitation of NS, which could in turn influence the macroscopic properties of NS\cite{joglekar2020relativistic, kouvaris2011constraining}.

Weakly interacting massive particles(WIMP), as one of the DM candidates, due to its freeze-out mechanism\cite{kouvaris2011constraining, baryakhtar2017dark}, WIMPs are the most abundant DM particles in the early Universe. Several studies calculated the NS properties with the inclusion of DM\cite{Bhat20,Quddus, Ivanytskyi20,Ellis18,panotopoulos2017dark,li2012too,leung2011dark,kouvaris2010can,sandin2009effects,ciarcelluti2011have, de2010neutron}. The central densities of NS differ by only about two orders of magnitude, from core to stellar surface, density spans nearly ten orders of magnitude, indicating that most of the mass is concentrated in the core. DeDeo and Psaltis\cite{dedeo2003towards} showed that the differences in the M-R relation of NS arising from different gravitational theories are significantly greater than those produced by different EOS within the same gravitational framework. Studying the influence of uncertainties in NM properties on the NS within a fixed gravitational theory remains meaningful. The extremely strong gravitational field of NS provides another perspective for constraining NM through spacetime curvature\cite{ekcsi2014does, rosi2015measurement}, redshift, and compactness.

In this work, based on the exp/emp ranges of $(E_{\text{sym}})_{n_0}$ and $L_{n_0}$, we investigate the effects of uncertainties on the properties of NM and the nonlinear $\sigma$-$\delta$ coupling together with the possible admixture of DM, on the macroscopic properties of NS. In Sec.\ref{sec:framework}, we present the theoretical framework, including the RMF formalism, the expressions for NM properties, the EOS of DM, and the mass-radius(M-R) relation, tidal deformability, curvature invariants of NS. In Sec.\ref{sec:result}, we present and discuss the numerical results. Finally, our conclusions are summarized in Sec.\ref{sec:summary}

\section{THEORETICAL FRAMEWORK}
\label{sec:framework}

In this section, we briefly sketch the RMF theory, Bayesian methods, the including of DM, TOV equations, tidal deformability, and curvature invariants.

\subsection{Relativistic Mean Field Theory}

The RMF Lagrangian consists of the interaction of mesons-nucleons, mesons self-couplings, cross-couplings, and the free-field of nucleons and mesons. We consider only the cubic and quartic self-couplings of the $\sigma$ meson to effectively control the incompressibility at $n_0$.
\begin{equation}
\begin{split}
    \mathcal{L}= & \frac{1}{2}\left(\partial_\mu \sigma \partial^\mu \sigma-m_\sigma^2 \sigma^2\right)-\frac{1}{4}W^{\mu \nu} W_{\mu \nu}+\frac{1}{2} m_\omega^2 \omega_\mu \omega^\mu \\ & -\frac{1}{4}\vec{R}^{\mu \nu} \cdot \vec{R}_{\mu \nu}+\frac{1}{2} m_{\boldsymbol{\rho}}^2 \boldsymbol{\rho}_\mu \cdot \boldsymbol{\rho}^\mu +\frac{1}{2} \partial_\mu \boldsymbol{\delta} \partial^\mu \boldsymbol{\delta}-\frac{1}{2} m_{\boldsymbol{\delta}}^2 \boldsymbol{\delta}^2 \\ &+\bar{\psi}\left(i \gamma^\mu \partial_\mu-m\right) \psi +g_\sigma \sigma \bar{\psi} \psi-g_\omega \omega_\mu \bar{\psi} \gamma^\mu \psi \\ &- \frac{1}{2} g_{\boldsymbol{\rho}} \boldsymbol{\rho}_\mu \cdot \bar{\psi} \gamma^\mu \boldsymbol{\tau} \psi+g_{\boldsymbol{\delta}} \boldsymbol{\delta} \cdot \bar{\psi} \boldsymbol{\tau} \psi-U(\sigma)
\end{split}
\label{eq:lagrangian}
\end{equation}
where $m_\sigma$, $m_\omega$, $m_{\boldsymbol{\rho}}$, and $m_{\boldsymbol{\delta}}$ are the meson masses, and $g_\sigma$, $g_\omega$, $g_{\boldsymbol{\rho}}$, $g_{\boldsymbol{\delta}}$ represent the coupling constants for nucleons with the corresponding mesons, $W^{\mu \nu}=\partial^\mu \omega^\nu-\partial^\nu \omega^\mu$ and $\vec{R}^{\mu \nu}=\partial^\mu \boldsymbol{\rho}^\nu-\partial^\nu \boldsymbol{\rho}^\mu$ are the field strength tensor of vector meson $\omega$, $\boldsymbol{\rho}$ respectively, the vector symbols represent isospin vectors, and $\boldsymbol{\tau}=\left(\tau_1, \tau_2, \tau_3\right)$ denote the 2$\times$2 Pauli isospin matrices. $U\left(\sigma\right)=\frac{1}{3} \bar{b} m\left(g_\sigma \sigma\right)^3+\frac{1}{4} \bar{c}\left(g_\sigma \sigma\right)^4$ is the self-coupling of $\sigma$, $\bar {b}$ and $\bar c$ are dimensionless coupling constants,
and $m$ is the vacuum mass of nucleus. By solving the Euler-Lagrange equation $\partial_\mu\left(\frac{\partial \mathcal{L}}{\partial\left(\partial_\mu \phi\right)}\right)-\frac{\partial \mathcal{L}}{\partial \phi}=0$, one can obtain the equations of motion for the nucleon and meson field. For simplicity, we use coupling constants with dimension $\text{MeV}^{-2}$, $C_\sigma^2=\frac{g_\sigma^2}{m_\sigma^2}, C_\omega^2=\frac{g_\omega^2}{m_\omega^2}, C_\rho^2=\frac{g_\rho^2}{m_\rho^2}, C_\delta^2=\frac{g_\delta^2}{m_\delta^2}$, the corresponding meson field becomes $\Phi=g_\sigma \sigma_0, \Omega=g_\omega \omega_0, \boldsymbol{b}=g_\rho \boldsymbol{\rho}_{0(3)}, \Delta=g_\delta \boldsymbol{\delta}_{0(3)}$, then the equations of motion given as
\begin{equation}
\Phi / C_\sigma^2+\bar{b} m \Phi^2+\bar{c} \Phi^3=n_{s p}+n_{s n}
\label{eq:sigma field}
\end{equation}
\begin{equation}
\Omega / C_\omega^2=n_p+n_n
\label{eq:omega field}
\end{equation}
\begin{equation}
\boldsymbol{b} / C_\rho^2=\frac{1}{2}\left(n_p-n_n\right)
\label{eq:rho field}
\end{equation}
\begin{equation}
\Delta / C_\delta^2=n_{s p}-n_{s n}
\label{eq:delta field}
\end{equation}
\begin{equation}
\left[\gamma^0 E_N^*-m_N^*\right] \psi_N(x)=0
\label{eq:dirac field}
\end{equation}
$E_{N}^*=\mu_N-g_\omega \omega_0-\frac{1}{2} g_\rho \tau_{3_N} \rho_{0(3)}=\sqrt{k_{F_N}^2+m_{\mathrm{N}}^{* 2}}$, $m_N^*=m-g_\sigma \sigma_0-g_\delta \tau_{3_N} \delta_{0(3)}$, where $\mu_N$ is nucleon chemical potential. N is taken as proton(p) or neutron(n). For protons, $\tau_{3p}=1$, neutrons, $\tau_{3n}=-1$. The subscript 0 indicates the time component, and for (3), the third component in the isospin space. $n_N$ and $n_ {sN}$ are vector density and scalar density of nucleon respectively, baryon number density $n_B = n_n + n_p$, $n_N = \frac {k_ {F_N} ^ 3} {3 \pi ^ 2}$. The scalar density $n_ {sN}= \frac{2}{(2 \pi)^3} \int_0^{k_{F_N}} \frac{m_N^*}{\sqrt{k^2+m_N^{* 2}}} d^3 k $ can be expressed in terms of the the nucleon effective energy $E_ {F_N} ^ \ast$ and Fermi momentum $k_{F_N}$ as
\begin{equation}
n_{sN}=\frac{m_N^*}{2 \pi^2}\left[E_{F_N}^* k_{F_N}-m_N^{* 2} \ln \frac{E_{F_N}^*+k_{F_N}}{m_N^*}\right]
\label{eq:scalar density}
\end{equation}
The EOS of NM is obtained through the energy-momentum tensor, energy is the time component $\varepsilon=\left\langle T^{00}\right\rangle$, pressure is the spatial component $P=\frac{1}{3}\left\langle T^{i i}\right\rangle$, and the energy-momentum tensor is given by
\begin{equation}
T^{\mu \nu}=\frac{\partial \mathcal{L}}{\partial\left(\partial_\mu \phi\right)} \partial^v \phi-g^{\mu \nu} \mathcal{L}
\end{equation}
the energy density $\varepsilon$ and the pressure P are
\begin{subequations}\label{eq:energy_pressure}
\begin{align}
\varepsilon= {}&\ \varepsilon_p^{k i n}+\varepsilon_n^{k i n}+\frac{1}{2} \frac{\Phi^2}{C_\sigma^2}+\frac{1}{3} \bar{b} m \Phi^3+\frac{1}{4} \bar{c} \Phi^4 \notag \\&+\frac{1}{2} \frac{\Omega^2}{C_\omega^2}+\frac{1}{2} \frac{\boldsymbol{b}^2}{C_\rho^2}+\frac{1}{2} \frac{\Delta^2}{C_\delta^2} \label{eq:energy} \\
P= {} &\ P_p^{kin}+P_n^{kin}-\frac{1}{2} \frac{\Phi^2}{C_\sigma^2}-\frac{1}{3} \bar{b} m \Phi^3-\frac{1}{4} \bar{c} \Phi^4 \notag \\& + \frac{1}{2} \frac{\Omega^2}{C_\omega^2}+\frac{1}{2} \frac{\boldsymbol{b}^2}{C_\rho^2}-\frac{1}{2} \frac{\Delta^2}{C_\delta^2} \label{eq:pressure}
\end{align}
\end{subequations}
$\varepsilon_N^{kin}$ and $P_N^ {kin }$ are kinetic contributions of the nucleons.
\begin{equation}
\varepsilon_N^{k i n}=\frac{1}{\pi^2} \int_0^{k_{F_N}} k^2 \sqrt{k+m_N^{* 2}} d k
\end{equation}
\begin{equation}
\mathrm{P}_N^{k i n}=\frac{1}{3 \pi^2} \int_0^{k_{F_N}} \frac{k^4}{\sqrt{k^2+m_N^{* 2}}} d k
\end{equation}

For NS matter, neutrons undergo $\beta$ decay into protons, electrons, and electron antineutrinos, $n \rightarrow p+e^{-}+\bar{\nu}_e$. To maintain charge neutrality, the inverse beta decay process $p+e^{-} \rightarrow n+\nu_e$ also occurs. In addition, muons can decay into electrons via $\mu^- \rightarrow e^- +\nu_\mu+\bar{\nu}_e$. To ensure the stability of NS matter, we consider leptonic degrees of freedom consisting of $e^-$ and $\mu^-$, while neglecting the contributions of neutrinos. Under these conditions, the chemical equilibrium and the charge neutrality can be expressed as
\begin{equation}
    \mu_n=\mu_p+\mu_e, \quad \mu_e=\mu_\mu
\end{equation}
\begin{equation}
    n_p=n_e+n_\mu
\end{equation}
the Lagrangian of lepton and its EOS can be written as
\begin{equation}
\mathcal{L}_l=\sum_l \bar{\psi}_l\left(i \gamma_\mu \partial^\mu-m_l\right) \psi_l
\label{eq:lagrangian lepton}
\end{equation}
\begin{subequations}
\begin{align}
&\varepsilon_l=\sum_{l=e, \mu} \frac{1}{\pi^2} \int_0^{k_F^l} \sqrt{k^2+m_l^2} k^2 d k \\ 
&P_l=\sum_{l=e, \mu} \frac{1}{3 \pi^2} \int_0^{k_F^l} \frac{k^4}{\sqrt{k^2+m_l^2}} d k
\end{align}
\end{subequations}
where $l$ takes $e^-$ and $\mu^-$, $m_l$ and $k_F^l$ denote the lepton mass and momentum respectively.

The intuitive interpretation of $E_{\text{sym}}$ is the energy required for SNM transformed into the pure neutron matter. Specifically, the EOS of neutron-rich matter can be expressed in terms of the binding energy per nucleon at the given baryon number density~($n_B$) within the parabolic approximation $E\left(n_B, \alpha\right)=E\left(n_B,\alpha=0\right)+E_{\mathrm{sym}}\left(n_B\right)\alpha^2+\mathcal{O}\left(\alpha^4\right)$, $\alpha=\left(n_n-n_p\right) /\left(n_p+n_n\right)$ is the neutron-proton asymmetry, $E\left(n_B, \alpha=0\right)$ is the average energy per nucleon of SNM and $E_{\text{sym}}\left(n_B\right)$ is the nuclear symmetry energy
\begin{equation}
    E_{\mathrm{sym}}\left(n_B\right)=\frac{1}{2}\left[\frac{\partial^2 E\left(n_B, \alpha\right)}{\partial \alpha^2}\right]_{\alpha=0}
\label{eq:symmetry energy}
\end{equation}

Typically, a few bulk parameters are used to characterize the behavior of SNM. Defining $E\left(n_B, \alpha=0\right)=E_0\left(n_B\right)$ and performing a Taylor expansion of the energy per nucleon and the symmetry energy around $n_0$
\begin{equation}
    E_0\left(n_B\right)=\varepsilon_0+\frac{1}{2} K x^2+\cdots
\end{equation}
\begin{equation}
    E_{\mathrm{sym}}\left(n_B\right)=J+L x+\frac{1}{2} K_{\mathrm{sym}} x^2+\cdots
\end{equation}
$\varepsilon_0$ and $K$ denote the energy per nucleon and the incompressibility of SNM at $n_0$. $J$, $L$, $K_{\text{sym}}$ represent the symmetry energy, its slope, and its curvature of NM at $n_B$, respectively. $x=\left(n_B - n_0\right) / 3 n_0$ is a dimensionless parameter that characterizes the deviation from the saturation density
\begin{equation}
    K=\left.9 n_0^2 \frac{d^2 E_0\left(n_B\right)}{d n_B^2}\right|_{n_0}
\label{eq:K0}
\end{equation}
\begin{equation}
    L=\left.3 n_0 \frac{d E_{\mathrm{sym}}\left(n_B\right)}{d n_B}\right|_{n_0}
\label{eq:Lsym}
\end{equation}
\begin{equation}
    K_{\text {sym }}=\left.9 n_0^2 \frac{d^2 E_{\text{sym}}\left(n_B\right)}{d n_B^2}\right|_{n_0}
\label{eq:skewness}
\end{equation}
the analytical expressions of the symmetry energy, incompressibility and symmetry energy slope at $n_0$\footnote{Throughout this work, the subscripts $0$ or $n_0$ denote quantities evaluated at the nuclear saturation density, including the symmetry energy $(E_{\text{sym}})_{n_0}=E_{\text{sym}}({n_0})$, incompressibility $K_0$, and symmetry energy slope $L_{n_0}$.} under the Lagrangian Eq.(\ref{eq:lagrangian}) are obtained by the definition Eq.(\ref{eq:symmetry energy})(\ref{eq:K0})(\ref{eq:Lsym})
\begin{widetext}
\begin{equation}
\begin{aligned}
    L_{n_0}=&\ \frac{k_{F_0}^2}{6E_{F_0}}\left[1+\frac{m_0^{\ast2}}{E_{F_0}^2}\left(1+\frac{3n_0C_\sigma^2}{E_{F_0}\left(1+C_\sigma^2A\right)}\right)\right]+\frac{3}{8}C_\rho^2n_0-\frac{C_\delta^2m_0^{\ast2}n_0}{2E_{F_0}^2\left(1+C_\delta^2A\right)}\left\{1+2\frac{m_0^{\ast2}}{E_{F_0}^2}\left[1+\frac{3n_0C_\sigma^2}{E_{F_0}\left(1+C_\sigma^2A\right)}\right] \right. \\ &\ \left. -\frac{3n_0C_\delta^2}{1+C_\delta^2A}\left[\frac{k_{F_0}^2}{E_{F_0}^3}-A_{m^\ast}^\prime\frac{m_0^\ast C_\sigma^2}{E_{F_0}\left(1+C_\sigma^2A\right)}\right]\right\}_{n_0}
\end{aligned}
\label{eq:Lsym no coupling}
\end{equation}
\end{widetext}
\begin{equation}
    E_{\text {sym }}(n_0)=\frac{k_0^2}{6 E_{F_0}}+\frac{1}{8} C_\rho^2 n_0-\frac{C_\delta^2 m_0^{* 2} n_0}{2 E_{F_0}^2\left(1+C_\delta^2 A\right)}
\label{eq:analytical Esym}
\end{equation}
\begin{equation}
\begin{aligned}
    K_0=9 n_0\left[\frac{\pi^2}{2 E_{F_0} k_0}+C_\omega^2-\frac{m_0^{* 2}}{E_{F_0}^2} \frac{C_\sigma^2}{1+C_\sigma^2 \frac{d^2 U(\sigma)}{d \Phi^2}+C_\sigma^2 A}\right]
\end{aligned}
\label{eq:analytical K0}
\end{equation}
where $A$ and $A_{m^*}^{\prime}$ are respectively the first- and second-order partial derivatives of the scalar density $n_s$ with respect to the effective mass of nucleon $m^*$
\begin{equation}
\begin{aligned}
    A & =\left. \frac{\partial n_s\left(m^*\right)}{\partial m^*}\right|_{n_0} \\ & =\frac{1}{\pi^2}\left[\frac{k_{F_0}}{E_{F_0}}\left(E_{F_0}^2+2 m_0^{* 2}\right)-3 m_0^{* 2} \ln \frac{k_{F_0}+E_{F_0}}{m_0^*}\right]
\end{aligned}
\end{equation}
\begin{equation}
\begin{aligned}
    A_{m^*}^{\prime} & =\left.\frac{\partial^2 n_s\left(m^*\right)}{\partial m^{* 2}}\right|_{n_0} \\ & =\frac{1}{\pi^2}\left[\frac{k_{F_0} m_0^*}{E_{F_0}^3}\left(3 E_{F_0}^2+k_{F_0}^2\right)-3 m_0^* \ln \frac{k_{F_0}+E_{F_0}}{m_0^2}\right]
\end{aligned}
\end{equation}

We represent the corresponding quantities at $n_0$ with $m_0^*$, $E_{F_0}$, $k_{F_0}$. Moreover, at zero temperature, the pressure can be transformed into Eq.(\ref{eq:pressure thermodynamic}) through thermodynamic relations, according to the pressure vanishes at $n_0$, a conclusion consistent with HVH theorem\cite{hugenholtz1958theorem} can be obtatined: the energy per nucleon equals to the Fermi energy
\begin{equation}
    P=-\left(\frac{\partial E}{\partial V}\right)_N=n_B \frac{\partial \varepsilon}{\partial n_B}-\varepsilon=n_B \left(\mu-\frac{E}{N}\right)
\label{eq:pressure thermodynamic}
\end{equation}
$E$ denotes the total energy, and $N$ here, the total nucleon number, which should be distinguished from the previous $N=p/n$. Eq.(\ref{eq:K0}) can be simplified as
\begin{equation}
K_0=\left.9 n_0^2 \frac{d^2 E_0\left(n_B\right)}{d n_{\mathrm{B}}^2}\right|_{n_0}=\left.9 n_0 \frac{d \mu}{d n_{\mathrm{B}}}\right|_{n_0}
\label{eq:K0 simplified}
\end{equation}

Although, the expressions Eq.(\ref{eq:analytical Esym})(\ref{eq:analytical K0})(\ref{eq:Lsym no coupling}) denote the result obtained by calculation at $n_0$, the incompressibility (\ref{eq:analytical K0}) is only applicable at $n_0$, the symmetry energy $E_{\text{sym}}(n_0)$ Eq.(\ref{eq:analytical Esym}) and the symmetry energy slope $L_{n_0}$~Eq.(\ref{eq:Lsym no coupling}) can be used at any density for SNM. Additionally, the Binding Energy(BE) 
\begin{equation}
BE=\varepsilon /n_B-m
\label{eq:BE}
\end{equation}

As described in the appendix of Ref.\cite{chen2014building}, $n_0$, BE, $m_0^*$, and $K_0$ can be calculated within isoscalar parameters, which correspond to $C_{\sigma}^2$, $C_{\omega}^2$, $\bar{b}$, $\bar{c}$ in this work. The saturation properties of NM are reasonably well constrained by experiments\cite{dutra2014relativistic}\cite{oertel2017equations}. A sensitivity analysis indicates that the quantities composed of isovector parameters $L$, $K_{\text{sym}}$ have a significant impact on the experimental parameters\cite{margueron2018equation}, also, values of $E_{\text{sym}}$ is obviously controlled by the isovector parameters.
Due to the large variation of the skewness $K_{\text{sym}}$ at $n_0$ according to different experiments and theoretical calculations, we chose $(E_{\text{sym}})_{n_0}$ and $L_{n_0}$ as coordinates to study NS within the range of exp/emp constraints on NM.

\subsection{Bayesian Inference}
\label{sec:bayes}

In the following, we will use Bayesian inference to constrain the isoscalar parameters under the current Lagrangian Eq.(\ref{eq:lagrangian}) and take it as the basis to discuss the distribution of the isovector parameters at $n_0$.

The principle of Bayesian inference is Bayes' theorem. By using the data of nuclear experiments, the model parameters required in theoretical calculations can be inferred. For RMF NM, the isoscalar parameters that need to be inferred are $C_{\sigma}^2$, $C_{\omega}^2$, $\bar b$, and $\bar c$.
The saturation properties of NM used to constrain are $n_0$, BE, $K_0$, and $m^*/m$. Particular, $n_0$ ranges from 0.14 to 0.162$\text{fm}^{-3}$, although $n_0$ is commonly taken to be around $0.16\text{fm}^{-3}$ in the literature,
many well-constrained RMF parameter sets, such as NL3\cite{lalazissis1997new}, G3\cite{kumar2017new}, FSUGold\cite{FSUGold}, and IOPB-I\cite{kumar2018new} etc., tend to predict a value closer to $0.15\text{fm}^{-3}$, we therefore take $0.151\text{fm}^{-3}$ as the center, and slightly extended the upper bound so that allow for the possibility of $0.16\text{fm}^{-3}$.
BE ranges from $-17$ to $-15$MeV\cite{huth2022constraining}, $K_0$ from 200 to 300MeV\cite{das2021impacts}\cite{huth2021new}\cite{stone2014incompressibility}, and $m^*/m$ from 0.55 to 0.85\cite{hornick2018relativistic}\cite{li2025influence}

Bayes' theorem states that the posterior distribution of the parameters $\boldsymbol{\theta}$ is proportional to the product of the corresponding prior distribution and the likelihood function.
\begin{equation}
p(\boldsymbol{\theta} \mid \boldsymbol{d}, \mathcal{M}) \propto p(\boldsymbol{\theta} \mid \mathcal{M}) p(\boldsymbol{d} \mid \boldsymbol{\theta}, \mathcal{M})
\end{equation}
in which $\boldsymbol{d}$ denotes the dataset, and $\mathcal{M}$ the model. $p(\boldsymbol{\theta} \mid \boldsymbol{d}, \mathcal{M})$ is the posterior probability distribution of $\boldsymbol{\theta}$, $p(\boldsymbol{\theta} \mid \mathcal{M})$ is the prior distribution of $\boldsymbol{\theta}$ for a given model, and $p(\boldsymbol{d} \mid \boldsymbol{\theta}, \mathcal{M})$, the likelihood function, which acts as the probability of the quantities for the given model and parameters.

In principle, the choice of the prior distribution for $\boldsymbol{\theta}$ can be arbitrary, as long as the posterior distribution converges to the true distribution under the constraints of the given dataset. However, due to practical computational limitations, it is necessary to specify a reasonably constrained range in order to save computational resources, with reference to the parameter ranges from various RMF models, Table.\ref{tab:prior distribution} presents the specific choices for the distributions of $\boldsymbol{\theta}$ used in this work.

\begin{table}[htbp]
\caption{The prior settings for isoscalar parameters. Here $\mathcal{U}$ denotes a uniform (flat) distribution.}
\begin{ruledtabular}
\begin{tabular}{cc}
isoscalar parameter & Prior \\
\hline
$C_{\sigma}^2$($\text{fm}^{2}$) & $\mathcal{U}(0,\,20)$ \\
$C_{\omega}^2$($\text{fm}^{2}$) & $\mathcal{U}(0,\,10)$ \\
$\bar b$ & $\mathcal{U}(-3,\,3)$ \\
$\bar c$ & $\mathcal{U}(-5,\,5)$ \\
\end{tabular}
\end{ruledtabular}
\label{tab:prior distribution}
\end{table}

We can rewrite the likelihood function with the assumption that all experiment measurements are independent
\begin{equation}
p(\boldsymbol{d} \mid \boldsymbol{\theta}, \mathcal{M}) = \prod_i p\left(X_i \mid d_{\exp / \text{emp}}, \mathcal{M}\right)
\end{equation}
$p\left(X_i \mid d_{\exp / \text{emp}},\mathcal{M}\right)$ denotes the probability that the physical quantity $i$ takes the value $X_i$ within the given exp/emp constraints and model, here $i$ takes $n_0$, BE, $K_0$ and $m^*/m$. The form of $p$ takes the Gaussian function, and the specific form is shown in Table.\ref{tab:Gaussian function}
\begin{table}[htbp]
\caption{Summary of the Gaussian funtion for $n_0$, BE, $K_0$ and $m^*/m$}
\begin{ruledtabular}
\begin{tabular}{cc}
$X_i$ & likelihood function \\
\hline
$n_0$ & $\mathcal{N}(0.151,\,0.0037^2)$ \\
$BE$ & $\mathcal{N}(-16,\,0.33^2)$ \\
$K$ & $\mathcal{N}(250,\,16.7^2)$ \\
$m^*/m$ & $\mathcal{N}(0.7,\,0.05^2)$ \\
\end{tabular}
\end{ruledtabular}
\label{tab:Gaussian function}
\end{table}

To ensure the convergence of the inference results, we performed two independent inference runs following the guidelines of emcee\cite{foreman2013emcee}\footnote{https://emcee.readthedocs.io/en/stable/}, and incorporated autocorrelation analysis during the process. Table.\ref{tab:comparison} presents two sets of parameters corresponding to the maximum posterior probability from each run, along with a comparison to the two sets of isoscalar parameters calculated in Kubis's work\cite{kubis1997nuclear}.

\begin{table*}[htbp]
\caption{The comparison of the maximum posteriori probability parameter obtained from two inferences with the parameters in Kubis\cite{kubis1997nuclear}.}
\begin{ruledtabular}
\begin{tabular}{lcccccccc}
 & $C_\sigma^2$ (fm$^2$) & $C_\omega^2$ (fm$^2$) & $\bar{b}$ & $\bar{c}$ & $m^*/m$ & $K_0$ (MeV) & $n_0$ (fm$^{-3}$) & BE (MeV) \\
\hline
parameters1 & 12.811 & 7.750 & 0.00371 & -0.00381 & 0.685 & 246.826 & 0.150 & -15.902 \\
parameters2 & 11.252 & 6.482 & 0.00368 & -0.000125 & 0.728 & 293.205 & 0.150 & -16.621 \\
Kubis stiff & 11.250 & 6.483 & 0.003825 & $3.5\times10^{-6}$ & 0.738 & 279.171 & 0.145 & -15.845 \\
Kubis soft & 1.582 & 1.019 & -0.7188 & 6.563 & 0.913 & 278.917 & 0.145 & -15.734 \\
\end{tabular}
\end{ruledtabular}
\label{tab:comparison}
\end{table*}

\begin{figure}
    \includegraphics[width=\linewidth]{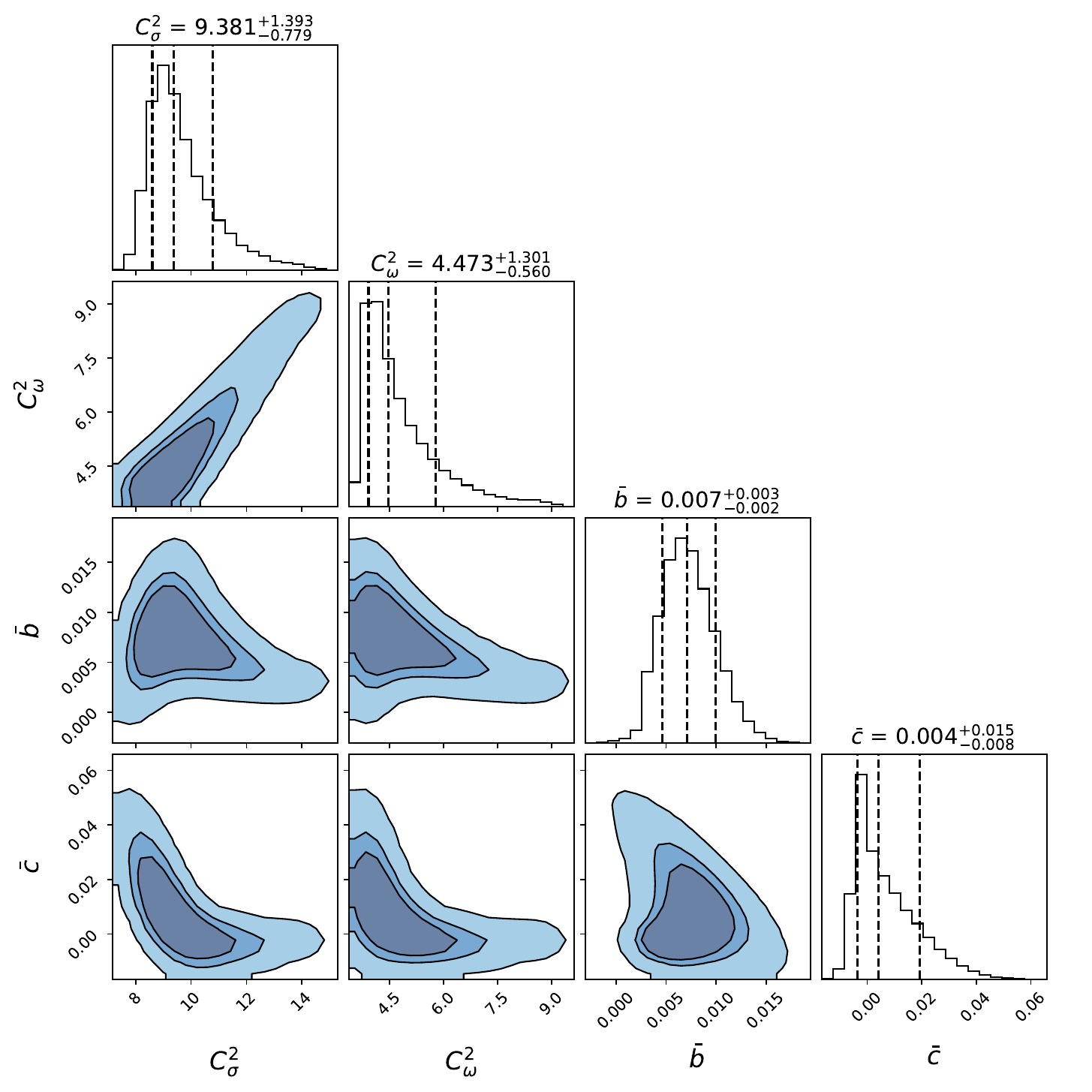}
    \caption{The posterior probability distribution of isoscalar parameters, with the diagonal being the posterior distribution of the parameters, the three vertical dotted lines represent the 16\%, 50\%, and 84\% quantiles respectively, the region between 16\% and 84\% is 1$\sigma$ credible region. The contour levels in the corner plot, going from dark blue to light blue correspond to the 68\%, 84\%, and 98.9\% credible region, respectively.}
    \label{fig:posterior}
\end{figure}

The saturation properties calculated using parameter sets 1 and 2 in Table.\ref{tab:comparison} show clear improvements compared to the two parameter sets in Kubis\cite{kubis1997nuclear}. In particular, $n_0$ in both cases is $0.15\text{fm}^{-3}$. Parameters1 significantly reduce the incompressibility from around 280MeV to 246MeV, while parameters2 is very close to the Kubis stiff. The posterior probability distribution(PPD) of parameters1 is shown in FIG.\ref{fig:posterior}.

It is evident that the PPD of the isoscalar parameters are all unimodal. For the non-negative, dimensionful coupling constants $C_\sigma^2$ and $C_\omega^2$, both their one-dimensional probability density distribution and two-dimensional joint PPD remain strictly positive, indicating that they are well constrained.
For the dimensionless $\sigma$ self-coupling constants $\bar{b}$ and $\bar{c}$, which mainly govern the incompressibility, are not subject to specific sign restrictions. Based on the above analysis, we take parameters1 as the foundation, the distribution of the isovector parameters $C_\rho^2$ and $C_\delta^2$ is explored.

\begin{figure}
    \includegraphics[width=\linewidth]{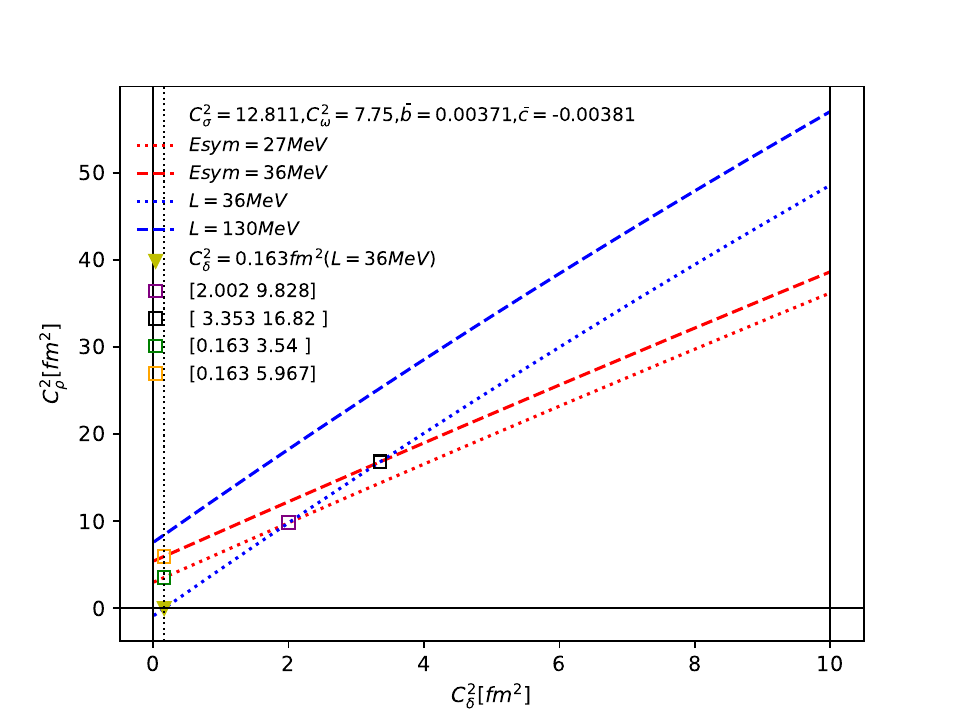}
    \caption{The variation of $C_\rho^2$ as a function of $C_\delta^2$ at $n_0$ evaluated at the upper and lower bounds of $E_{\text{sym}}$ and $L$. The red curve indicated the upper and lower bounds of the $E_{\text{sym}}$, the blue curve represents $L$, the vertical black dotted line at $C_\delta^2 = 0.163 \text{fm}^2$ marks the physically allowed region where $C_\rho^2$ remains positive. The quadrilateral enclosed by the four open square markers corresponds to the allowed range of the isovector parameters within the bounds of $E_{\text{sym}}$ and $L$.}
    \label{fig:isovector}
\end{figure}

For $E_{\text{sym}}$ and $L$, we adopt the ranges $(E_{\text{sym}})_{n_0} \in [27,~36]$~MeV, $L_{n_0} \in [36,~130]$~MeV\cite{you2025u}\cite{xia2022unified}. In FIG.\ref{fig:isovector}, we present the range of $C_\rho^2$ as a function of $C_\delta^2$ at $n_0$ within the limits of $(E_\text{sym})_{n_0}$ and $L_{n_0}$.

As shown in FIG.\ref{fig:isovector}, the red and blue curves correspond to $C_\rho^2$ evaluated at the boundary values of $(E_\text{sym})_{n_0}$ and $L_{n_0}$, respectively. The four open square markers denote the intersections associated with different combinations of $(E_\text{sym})_{n_0}$ and $L_{n_0}$. In particular, the orange and green squares mark the intersections between the two boundary curves of $(E_\text{sym})_{n_0}$ and the line $C_\delta^2=0.163$. Hereafter, for convenience in expression, when the behavior characteristics are no longer stressed, both $E_{\text{sym}}$ and $L$ represent the values at $n_0$.

$C_\delta^2=0.163$ serves as a simple constraint to ensure that the $C_\rho^2$ remains positive; it corresponds to the value of $C_\delta^2$ for which $C_\rho^2=0$ at $L=36 \mathrm{MeV}$. Although physical acceptable parameter sets may still exist for $C_\delta^2<0.163$, this condition is adopted as a lower bound on $C_\delta^2$ to avoid potential unphysical behavior. The quadrilateral enclosed by these four open squares 
therefore defines the region of physically reasonable values for $C_\rho^2$ and $C_\delta^2$.

Taking the upper and lower red curves in FIG.\ref{fig:isovector} as an example, $\boldsymbol{(a).}$Considering the blue dotted line corresponding to $L=36~\text{MeV}$, as it moves from the lower red dotted curve($E_{\mathrm{sym}}=27 ~\mathrm{MeV}$) to the upper one($E_{\mathrm{sym}}=36 ~\mathrm{MeV}$), $L$ is fixed at 36~MeV, while $E_{\text{sym}}$ increases from 27 to 36~MeV. Similarly, $\boldsymbol{(b).}$Between the two boundary curves of $L$, there exist trajectories along which $L$ varies continuously while $E_{\text{sym}}$ remains constant.

Although in FIG.\ref{fig:isovector} the upper bound $L=130\text{MeV}$ is sufficiently large that it does not intersect with either boundary of $E_{\text{sym}}$, it is expected that the intersections will occur once $L$ decreases to smaller values.

It should be emphasized that, when considering the case of fixed $E_{\text{sym}}$ with continuously varying $L$, and scanning $C_\delta^2$ from smaller to larger values, the $E_{\text{sym}}$ fixed curves intersect the upper boundary of $L$ first. Consequently, for the same process of increasing $C_\delta^2$, the first type of behavior $\boldsymbol{(a).}$ corresponds to fixed $L$ with increasing $E_{\text{sym}}$, whereas the second, $\boldsymbol{(b).}$ corresponds to fixed $E_{\text{sym}}$ with decreasing $L$. Hence the uncertainties in the coupling constants are mapped onto variations in $E_{\text{sym}}$ and $L$.

The orange and green points in the plane $\left(C_\delta^2, C_\rho^2\right)$ are given by (0.163, 5.967)$\text{fm}^2$ and (0.163, 3.540)$\text{fm}^2$, respectively, corresponding to $\left(E_{\text{sym}}, L\right)=(36.0, 102.4)$MeV and $\left(E_{\text{sym}}, L\right)=(27.0, 75.4)$MeV. To ensure that the curve $C_\rho^2$ with fixed $L$ fully traverses the region bounded by the upper and lower limits of $E_{\text{sym}}$ within the quadrilateral, the maximum value of $L$ should be lower than that of the green square, $L=75.4$~MeV. Furthermore, in order to maintain comparable variation intervals in the two cases discussed above, and taking into account the constraints adopted in the previous discussion, the range of $L$ finally chosen to be 40 $\sim$ 70MeV\cite{li2013constraining}\cite{zhu2018neutron}.

\subsection{$\sigma$-$\delta$ coupling}

The inclusion of the $\sigma-\delta$ coupling softens the symmetry energy in intermediate-density while stiffening it at high densities, thus alleviating the tension between the constraint of GW170817 and those inferred from nuclear experiments\cite{170817}\cite{danielewicz2002determination}\cite{zabari2019influence}. We consider a nonlinear $\sigma-\delta$ coupling. The corresponding coupling constant $g_{\sigma \delta}$ is taken to be -0.004, following Zabari et al\cite{zabari2019influence}. The modifications to the field equations, the EOS, and the expressions for nuclear matter properties arising from the inclusion of this coupling are presented below. For simplicity, here both the $\sigma$ and the $\delta$ fields are written directly in the form of the mean field($\Phi$ and $\Delta$).

The Lagrangian Eq.(\ref{eq:lagrangian}) becomes

\begin{equation}
\mathcal{L} \rightarrow \mathcal{L}^{\prime}=\mathcal{L}+\mathcal{L}_{\sigma \delta}
\label{eq:lagrangian prime}
\end{equation}
where $\mathcal{L}_{\sigma \delta}=-g_{\sigma \delta} \Phi^2 \Delta^2$, the $\sigma$ and $\delta$ field equations become
\begin{equation}
\Phi / C_\sigma^2+\bar{b} m \Phi^2+\bar{c} \Phi^3+2 g_{\sigma \delta} \Phi \Delta^2=n_{s p}+n_{s n}
\end{equation}
\begin{equation}
\Delta / C_\delta^2+2 g_{\sigma \delta} \Delta \Phi^2=n_{s p}-n_{s n}
\end{equation} 
based on the previous foundation (\ref{eq:energy}) and (\ref{eq:pressure}), the EOS becomes
\begin{subequations}\label{eq:energy_pressure prime}
\begin{align}
& \varepsilon ~\rightarrow ~ \varepsilon^{\prime} = ~ \varepsilon +g_{\sigma \delta} \Phi^2 \Delta^2 \label{eq:energy prime}\\
& P \rightarrow P^{\prime}=P-g_{\sigma \delta} \Phi^2 \Delta^2 \label{eq:pressure prime}
\end{align}
\end{subequations}
Similarly, $E_{\text{sym}}$ and $L$ can be written as
\begin{equation}
E_{\text {sym}}(n)=\frac{k_0^2}{6 E_{F_0}}+\frac{1}{8} C_\rho^2 n_0-\frac{C_\delta^2 m_0^{* 2} n_0}{2 E_{F_0}^2\left(1+C_\delta^2 A+g_{\sigma \delta} C_\delta^2 \Phi^2\right)}
\label{eq:Esym coupling}
\end{equation}

\begin{widetext}
\begin{equation}
\begin{aligned}
    L=&\ \frac{k_{F_0}^2}{6E_{F_0}}\left[1+\frac{m_0^{\ast2}}{E_{F_0}^2}\left(1+\frac{3n_0C_\sigma^2}{E_{F_0}\left(1+C_\sigma^2A\right)}\right)\right]+\frac{3}{8}C_\rho^2n_0-\frac{C_\delta^2m_0^{\ast2}n_0}{2E_{F_0}^2\left(1+C_\delta^2A+g_{\sigma\delta}C_\delta^2\Phi^2\right)}\left\{1+2\frac{m_0^{\ast2}}{E_{F_0}^2}\left[1+\frac{3n_0C_\sigma^2}{E_{F_0}\left(1+C_\sigma^2A\right)}\right] \right. \\ &\ \left. -\frac{3n_0C_\delta^2}{1+C_\delta^2A+g_{\sigma\delta}C_\delta^2\Phi^2}\left[\left(4g_{\sigma\delta}\Phi-A_{m^\ast}^\prime\right)\frac{m_0^\ast C_\sigma^2}{E_{F_0}\left(1+C_\sigma^2A\right)}+\frac{k_{F_0}^2}{E_{F_0}^3}\right]\right\}_{n_0}
\end{aligned}
\label{eq:Lsym coupling}
\end{equation}
\end{widetext}

\begin{figure}
    \includegraphics[width=\linewidth]{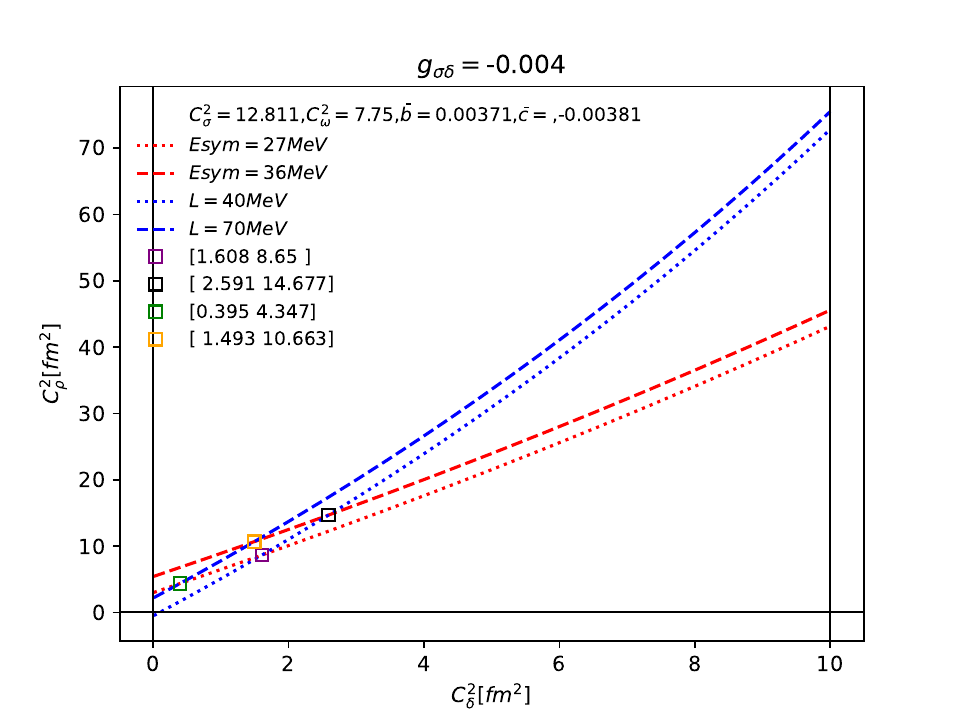}
    \caption{The variation of $C_\rho^2$ as a function of $C_\delta^2$. Same as FIG.\ref{fig:isovector}, except the $g_{\sigma \delta}=-0.004$ included and parameter1 from TAB.\ref{tab:comparison} is adopted. As discussed in SEC.\ref{sec:bayes} the allowed range of $L_{n_0}$ is restricted to $40\sim70$~MeV.}
    \label{fig:isovector_coupling}
\end{figure}

In FIG.\ref{fig:isovector_coupling}, we present the variation of $C_\rho^2$ as a function of $C_\delta^2$ including the $\sigma-\delta$ coupling, based on the parameters1 in TAB.\ref{tab:comparison}, in the same format as FIG.\ref{fig:isovector}. A slight modification has been made by adjusting the upper and lower bounds of $L$ to 70 and 40MeV, in the subsequent calculations of the RMF EOS, both with and without the $\sigma-\delta$ coupling, we perform the same calculations within this range.

\subsection{Interaction Between Nuclear Matter and DM}

We consider fermionic dark matter(DM) at fixed Fermi momentum, which interacts indirectly with nucleons via the Standard Model(SM) Higgs boson. The corresponding Lagrangian is given by\cite{panotopoulos2017dark}
\begin{equation}
\begin{aligned}
    \mathcal{L}_{\mathrm{DM}} = & \bar{\chi}\left[i \gamma^\mu \partial_\mu-M_\chi+y h\right] \chi+\frac{1}{2} \partial_\mu h \partial^\mu h \\ &\ -\frac{1}{2} M_h^2 h^2+f \frac{m}{v} \bar{\psi} h \psi
\end{aligned}
\end{equation}
the lightest neutralino $M_\chi=200 \mathrm{GeV}$ which acts as a fermionic dark matter candidate is considered\cite{panotopoulos2017dark}\cite{murakami2001nucleon}. $\chi$ is fermionic dark matter, here denotes the corresponding dark Fermion field, and $y=0.07$\cite{panotopoulos2017dark}\cite{murakami2001nucleon} is the coupling between the DM and the Higgs field $h$. The interaction between $h$ and nucleons occurs through the effective Yukawa coupling $\frac{fm}{v}$, $v=246$GeV is the vacuum expectation values of Higgs, the mass is $M_h=125$GeV and $f=0.35$\cite{cline2013update}\cite{das2020effects} is the proton-Higgs form factor. The terms $h^3$ and $h^4$ in Higgs potential are not considered, because in the solution of the mean field, the value of $h$ is approximately $10^{-8}$ to $10^{-10}$ of other meson fields. Defining $\mathcal{L}_{\mathrm{NM}}$ is the Lagrangian of nuclear matter, compared to the field equations already given in the previous RMF theory, the additional modified field equations under the total Lagrangian $\mathcal{L}_{\text {total }}=\mathcal{L}_{\text {NM}}+\mathcal{L}_{\mathrm{DM}}$ here are
\begin{equation}
M_h^2 h_0=y\langle\bar{\chi} \chi\rangle+f \frac{m}{v}\langle\bar{\psi} \psi\rangle
\label{eq:higgs field equation}
\end{equation}
\begin{equation}
\left(i \gamma^\mu \partial_\mu-M_\chi^{\star}\right) \chi=0
\label{eq:DM dirac}
\end{equation}
\begin{equation}
    \left[\gamma^0 E_N^*-m_N^*\right] \psi_N(x)=0
\label{eq:NM dirac DM }
\end{equation}
where $\langle\bar{\psi} \psi\rangle$ is the nucleon scalar density(\ref{eq:scalar density}), the effective mass of the nucleons and the DM particles given as
\begin{equation}
    m_N^*=m-g_\sigma \sigma_0-\tau_{3_N} g_\delta \delta_{0(3)} -f \frac{m}{v} h_0
\label{eq:nucleon eff mass H}
\end{equation}
\begin{equation}
    M_\chi^*=M_\chi-y h_0
\label{DM eff mass}
\end{equation}
$\langle\bar{\chi} \chi\rangle$ and $k_F^{\text{DM}}$ are DM scalar density$(n_s^{\mathrm{DM}})$ and Fermi momentum, respectively.
\begin{equation}
  n_s^{\text{DM}}=\frac{2}{(2 \pi)^3} \int_0^{k_F^{\text{DM}}} \frac{M_\chi^{\star}}{\sqrt{k^2+M_\chi^{*2}}} d^3 k
\end{equation}
EOS for DM
\begin{subequations}
\begin{align}
&\varepsilon_{\text{DM}} = \frac{1}{\pi^2} \int_0^{k_F^{\mathrm{DM}}} k^2 \sqrt{k^2+\left(M_\chi^*\right)^2} d k+\frac{1}{2} M_h^2 h_0^2 \\
&P_{\mathrm{DM}}=\frac{1}{3 \pi^2} \int_0^{k_F^{\mathrm{DM}}} \frac{k^4}{\sqrt{k^2+\left(M_\chi^*\right)^2}} d k-\frac{1}{2} M_h^2 h_0^2
\end{align}
\end{subequations}

\subsection{TOV equation, and Tidal Deformability}
The mass-radius relation of the compact star can be obtained using the EOS and hydrostatic equilibrium equations, in General Relativity(GR), the equilibrium of a static, spherically symmetric compact star is described by the Tolman-Oppenheimer-Volkoff (TOV) equations\cite{tolman1939static}\cite{oppenheimer1939massive}
\begin{equation}
\frac{d P(r)}{d r} = -\frac{G M \varepsilon}{r^2} \frac{(1+P / \varepsilon)\left(1+4 \pi r^3 P / M\right)}{1-2 G M / r}
\label{eq:TOV1}
\end{equation}
\begin{equation}
\frac{d M(r)}{d r} =4 \pi r^2 \varepsilon(r)
\label{eq:TOV2}
\end{equation}
where $P(r)$ and $\varepsilon(r)$ are the pressure and energy density at the radius $r$, respectively, and $M(r)$ is the gravitational mass enclosed within $r$. $G=6.707 \times10^{-45} \mathrm{MeV}^{-2}$ denotes the gravitational constant. To solve the TOV equations, we need boundary conditions at the center and the surface of the stellar, $\varepsilon(0)=\varepsilon_{\mathrm{c}}$, $M(0)=0$, $P\left(r^*\right)=0$, $r^*$ is the radius of the stellar surface.

In addition to homogeneous nuclear matter(RMF) in the core of NS, the internal structure also includes non-uniform inner and outer crust. The inner crust is expected to be nonuniform neutron-rich matter, where the nuclei undergo deformation due to the large pressure, which may lead to the arising of pasta phases composed of nuclei with different geometrical configurations\cite{xia2022unified}\cite{bao2015impact}\cite{carriere2003low}. 
We use a polytropic EOS to avoid the complexity of the pasta structure with $P(\varepsilon)=A+B \varepsilon^{4 / 3}$, $A$ and $B$ are the coefficients determined by the EOS endpoints of the core and outer crust layers, BPS EOS\cite{baym1971ground} used for the outer crust, details for the EOS matching refer to Ref.\cite{fortin2016neutron}. In this work, the matching region for polytropic EOS was chosen from 0.5$n_0$ to $n_0$. 

The tidal deformability($\Lambda$) is a dynamical property of matter subject to a tidal field, its effects in binary systems has been focus of intense research in recent years\cite{flanagan2008constraining, Hinderer:2007mb, Damour:2009vw, malik2018gw170817}.
Studying $\Lambda$ requires solving the TOV equations simultaneously, since the stellar mass and radius entering the calculation are not independent inputs but are determined self-consistently during the integration of the TOV equations. The tidal Love number $k_2$ is given by\cite{Hinderer:2007mb}
\begin{equation}
\begin{aligned}
k_2=& \frac{8 \mathcal{C}^5}{5}(1-2 \mathcal{C})^2\left[2-y_R+2 \mathcal{C}\left(y_R-1\right)\right] \\ &\ \times \left\{2 \mathcal{C}\left[6-3 y_R+3 \mathcal{C}\left(5 y_R-8\right)\right]\right. \\ &\ +4 \mathcal{C}^3\left[13-11 y_R+\mathcal{C}\left(3 y_R-2\right)+2 \mathcal{C}^2\left(1+y_R\right)\right] \\&\ \left.+3(1-2 \mathcal{C})^2\left[2-y_R+2 \mathcal{C}\left(y_R-1\right)\right] \log (1-2 \mathcal{C})\right\}^{-1}
\end{aligned}
\label{eq:k2}
\end{equation}
where $\mathcal{C} = M / r$ is the dimensionless compactness parameter, $y_R \equiv y(r)$ is the solution of the following differential equation\cite{Postnikov:2010yn}
\begin{equation}
r \frac{d y(r)}{d r}+y(r)^2+y(r) F(r)+r^2 Q(r)=0
\label{eq:yr}
\end{equation}
$F(r)$ and $Q(r)$ are
\begin{equation}
    F(r)=\left[1-4 \pi r^2(\varepsilon-P)\right]\left[1-\frac{2 M}{r}\right]^{-1}
\end{equation}
\begin{equation}
\begin{aligned}
    Q(r)=&\ 4 \pi\left[5 \varepsilon+9 P+\frac{\varepsilon+P}{c_s^2}-\frac{6}{4 \pi r^2}\right]\left[1-\frac{2 M}{r}\right]^{-1} \\ &\ -\frac{4 M^2}{r^4}\left[1+\frac{4 \pi r^3 P}{M}\right]^2\left[1-\frac{2 M}{r}\right]^{-2}
\end{aligned}
\end{equation}
where $c_s^2(r)=\partial P(r) / \partial \varepsilon(r)$ is the squared speed of sound. The stellar surface is defined by the vanishing of pressure $P(r^*)=0$. Across the interface from the surface to the vacuum, a discontinuity in the energy density arises, the corresponding jump condition for $y_R$ is\cite{Damour:2009vw}\cite{Postnikov:2010yn}\cite{Zhou:2017pha}
\begin{equation}
y_R \rightarrow y_R-\frac{4 \pi r^3 \varepsilon_s}{M},
\end{equation}
$\varepsilon_s$ is energy density difference between the stellar surface and the vacuum. The boundary condition for Eq.(\ref{eq:yr}) at $r=0$ is given by $y(0)=2$. After obtaining $y_R$, the tidal Love number $k_2$ can be determined. $\Lambda$ can then be calculated from the following expression
\begin{equation}
\Lambda=\frac{2}{3} k_2 r^5
\end{equation}

\subsection{Curvature}
Curvature provides a fundamental measure for quantifying gravity and characterizing the strength of the gravitational field\cite{Psaltis:2008bb}. In the study of general relativity and NS, four types of curvature are commonly used to describe the spacetime structure both inside and outside the stars, the  Ricci scalar $\mathcal{R}$, the full contraction of the Ricci tensor $\mathcal{J}$, the full contraction of the Riemann tensor $\mathcal{K}$, and the full contraction of the Weyl tensor $\mathcal{W}$. We adopt the mathematical form of these curvature invariants and the compactness($\eta$) for spherically symmetric metric in GR presented in Ref.\cite{ekcsi2014does}
\begin{equation}
\mathcal{R}(r)=\kappa\left(\varepsilon c^2-3 P\right), \quad \kappa \equiv \frac{8 \pi G}{c^4}
\label{eq:Ricci scalar}
\end{equation}
\begin{equation}
\mathcal{J}^2 \equiv \mathcal{R}_{\mu \nu} \mathcal{R}^{\mu \nu}=\kappa^2\left[\left(\varepsilon c^2\right)^2+3 P^2\right]
\label{eq:Ricci tensor}
\end{equation}
\begin{equation}
\begin{aligned}
\mathcal{K}^2  &\ \equiv \mathcal{R}^{\mu \nu \rho \sigma} \mathcal{R}_{\mu \nu \rho \sigma} \\ &\  =\kappa^2\left[3\left(\varepsilon c^2\right)^2+3 P^2 +2 P \varepsilon c^2\right]\\ &\ ~~~ -\frac{16 \kappa G M \varepsilon}{r^3} +\frac{48 G^2 M^2}{r^6 c^4}
\label{eq:Kretschmann scalar}
\end{aligned}
\end{equation}
\begin{equation}
~~~~~~~~~~~~\mathcal{W}^2 \equiv \mathcal{C}^{\mu \nu \rho \sigma} \mathcal{C}_{\mu \nu \rho \sigma}=\frac{4}{3}\left(\frac{6 G M}{c^2 r^3}-\kappa \varepsilon c^2\right)^2
\label{eq:Weyl tensor}
\end{equation}
\begin{equation}
\eta(r) \equiv \frac{2 G M}{r c^2}
\label{eq:compactness}
\end{equation}
where $\varepsilon$ and $P$ denote the energy density and the pressure at $r$, respectively, and $M$ denotes the enclosed stellar mass within $r$. 

From above expressions, it is evident that $\mathcal{R}$ and $\mathcal{J}$ vanish outside the stellar, as they depend on the NS EOS. In contrast, the full contraction of the Riemann tensor also called Kretschmann scalar, it has nonvanishing components in vacuum, e.g. $\mathcal{R}^1{ }_{010}=-2 G M / c^2 r^3$, and is therefore more suitable than $\mathcal{R}$ and $\mathcal{J}$ for characterizing the curvature of the spacetime. Moreover, the square root of the full contraction of the Weyl tensor coincides with that of the Kretschmann scalar in vacuum. Consequently, $\mathcal{K}$ and $\mathcal{W}$ provide suitable measures of the spacetime curvature both inside and outside the NS\cite{ekcsi2014does}. These curvatures take different values inside the stellar, whereas $\mathcal{K}$ and $\mathcal{W}$ become identical in vacuum.

\section{RESULTS AND DISCUSSIONS}
\label{sec:result}

In this section, we present the impact of changes in NM properties($(E_{\text{sym}})_{n_0}$ and $L_{n_0}$) on the $M-R$ relation, the $M-\Lambda$ relation, and the curvature invariants.

\subsection{Stiffness of $E_{\text{sym}}$}
\label{sec:Esym}

\begin{figure}
    \includegraphics[width=\linewidth]{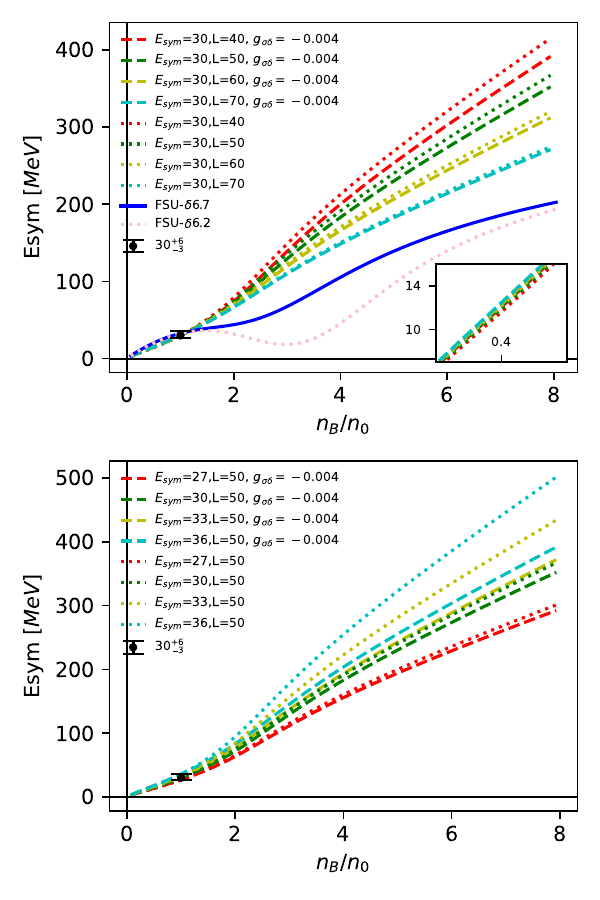}
    \caption{The density dependence of $E_{\text{sym}}$ with different combinations of $(E_{\text{sym}})_{n_0}$ and $L_{n_0}$ for cases with and without the $\sigma-\delta$ coupling. In the first set of panels, $(E_{\text{sym}})_{n_0}$ is fixed at 30~MeV, $(L)_{n_0}=40,50,60,70$~MeV. the blue solid line and the pink dotted line represent the behavior of $E_{\text{sym}}$ for the FSU-$\delta 6.7$ and FSU-$\delta 6.2$ parametrizations from \cite{Li:2022okx}, respectively. In the second, $L_{n_0}$ is fixed at 50~MeV, $(E_{\text{sym}})_{n_0}=27,30,33,36$~MeV. In the upper panel, the inset in the lower-right corner shows the behavior in the range $n_B/{n_0} = 0.2\sim0.6$.}
    \label{fig:EsymLcomparison1}
\end{figure}

In the present work, we employ Bayesian inference to obtain a set of parameters consistent with the current exp/emp constraints. The corresponding saturation properties are summarized in Table.\ref{tab:comparison}. FIG.\ref{fig:EsymLcomparison1} displays the density dependence of $E_{\text{sym}}$ for two representative cases. In the first case, $(E_{\text{sym}})_{n_0}$ is fixed at 30MeV, while $L_{n_0} = 40, 50, 60,70$MeV. In the second case, $L_{n_0}$ is fixed at 50MeV, $(E_{\text{sym}})_{n_0}=27,30,33,36$MeV. These results illustrate the evolution of $E_{\text{sym}}$ with $n_B$ under different conditions at $n_0$.
Compared with the results for $\text{FSU}-\delta 6.7$ and $\text{FSU}-\delta 6.2$ in Li \textit{et al.}\cite{Li:2022okx}, the differences stem from the fact that the Lagrangian Eq.(\ref{eq:lagrangian}) and Eq.(\ref{eq:lagrangian prime}) does not include the $\omega$--$\rho$ coupling and the self-interaction of the $\omega$ meson. Nevertheless, the general trend is preserved: $E_{\mathrm{sym}}$ exhibits softening behavior at intermediate densities and becomes stiffer at higher densities~\cite{zabari2019influence}.

The density-dependent behavior of $E_{\mathrm{sym}}$ both with and without the $\sigma$--$\delta$ coupling, has a clear tendency around $n_0$. Below $n_0$, larger values of $L_{n_0}$ correspond to a stiffer behavior of $E_{\text{sym}}$, however, above $n_0$, $E_{\text{sym}}$ shows a softer behavior, thus a reversal is observed in the stiffness ordering of $E_{\text{sym}}$. In particular, the curves of $E_{\text{sym}}$ that are softer below $n_0$ can become stiffer once the density exceeds $n_0$. We call this phenomenon \textbf{stiffness crossing} to describe the subsequent relevant phenomenon for simplicity. Similarly, $\sigma-\delta$ coupling with $(E_{\text{sym}})_{n_0}$ fixed stiffens $E_{\text{sym}}$ below $n_0$, while it softens $E_{\text{sym}}$ above $n_0$. 

With $L_{n_0}$ fixed at 50MeV, $E_{\text{sym}}$ shows a softer hehavior with smaller $(E_{\text{sym}})_{n_0}$. in this case, the crossing behavior is not observed. Compared to the scenario without the $\sigma-\delta$ coupling, $E_{\text{sym}}$ is consistently softer when the coupling is included.

Notably, the ordering of the softening-stiffening behavior of the $E_{\text{sym}}$ curves for different $L_{n_0}$ around $n_0$ is not a fixed conclusion, but varies with the adopted isoscalar parameters. A comparison with the results of Zhu et al.\cite{zhu2018neutron} illustrates this point, which will be analyzed in detail in the following section.

\subsection{$M-R-\Lambda$ Relation}
\label{sec:MRLambda relation}

\begin{figure*}[t]
\centering
\includegraphics[width=\textwidth]{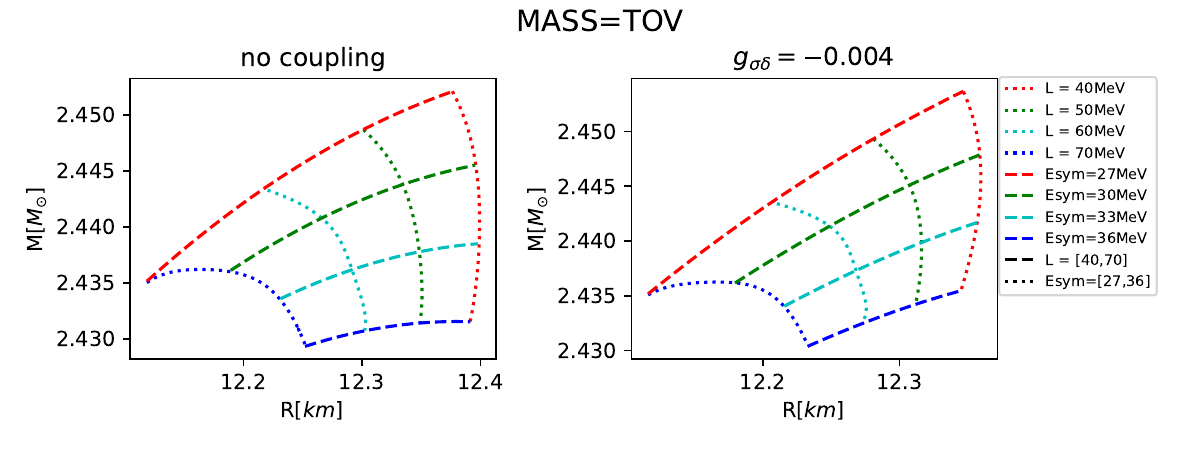}
\caption{$M-R$ relations of TOV configurations with and without the $\sigma-\delta$ coupling. In the legend, the black dotted and dashed lines represent the continuous variation ranges of $E_{\text{sym}}$ and $L$ at $n_0$. The colored dashed lines correspond to the fixed values of $(E_{\text{sym}})_{n_0}=27, 30, 33, 36$~MeV, with $L_{n_0}$ varying within range $[40, 70]$~MeV. Another case, the dotted lines denote the fixed values of $L_{n_0}=40, 50, 60, 70$~MeV, while $(E_{\text{sym}})_{n_0}$ varies in $[27, 36]$~MeV.}
\label{fig:MR comparison}
\end{figure*}

\begin{figure*}[t]
\centering
\includegraphics[width=\textwidth]{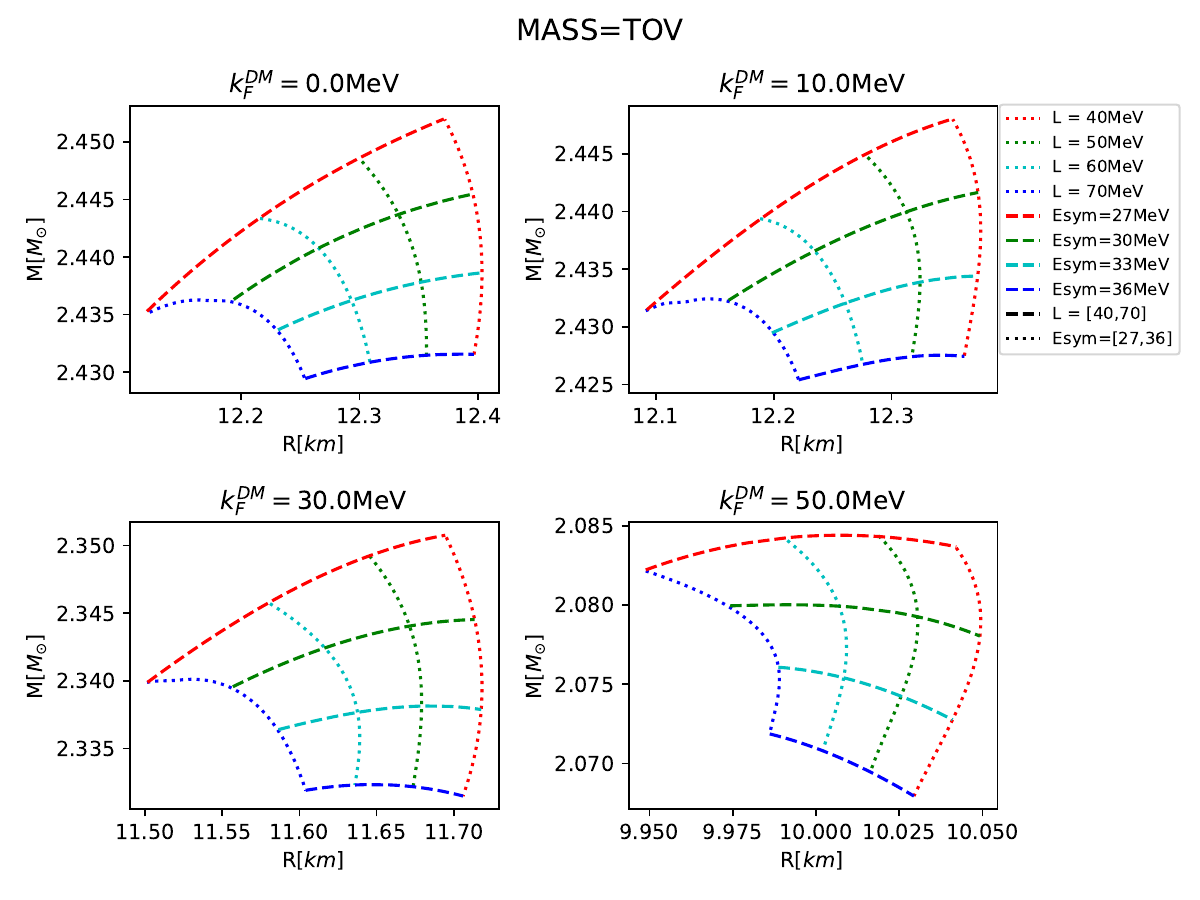}
\caption{$M-R$ relations of the TOV configurations with including of DM, The legend follows the same convention as in FIG.\ref{fig:MR comparison}. With increasing $k_F^{\text{DM}}$, the EOS undergoes stronger softening, resulting in a progressively smaller slope of the $M-R$ relation as shown in dashed lines.}
\label{fig:DM MR comparison}
\end{figure*}

\begin{figure*}[t]
\centering
\includegraphics[width=\textwidth]{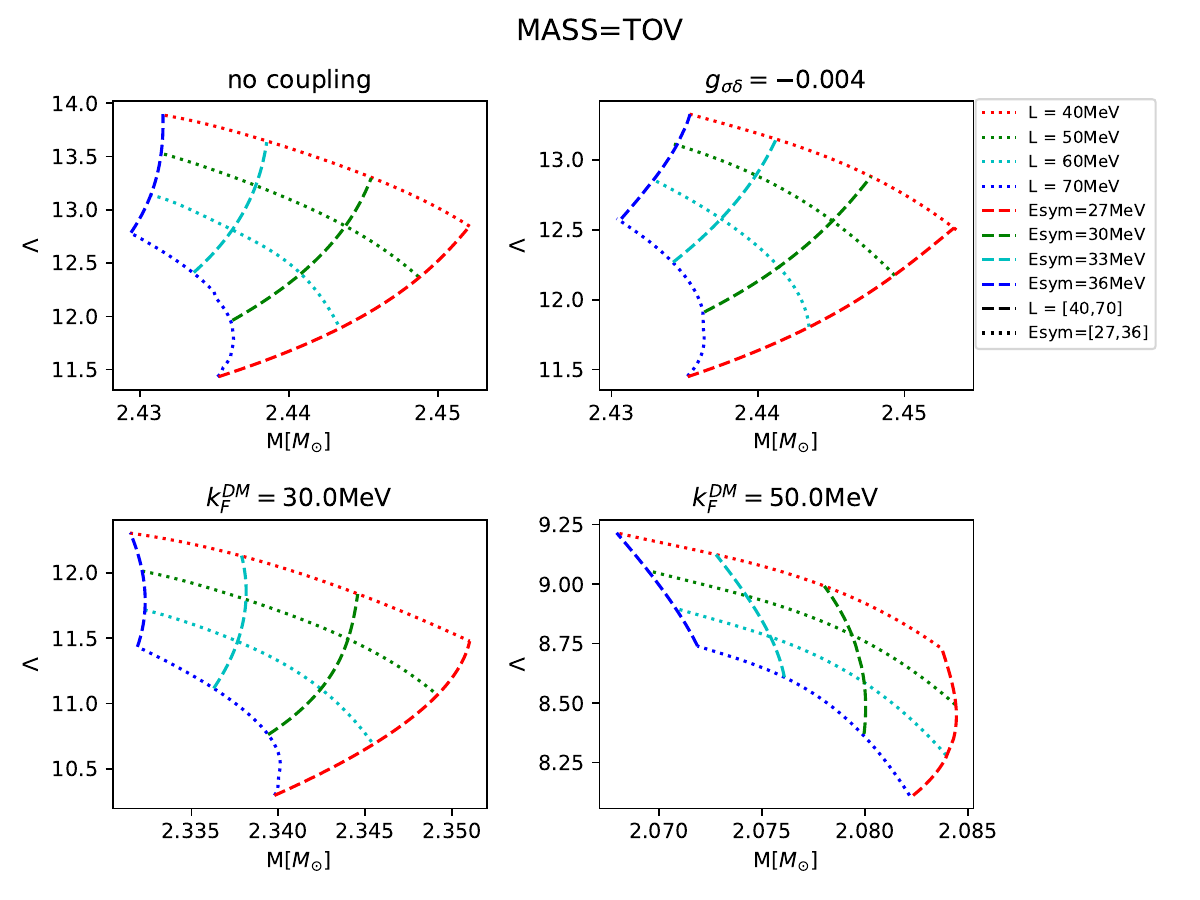}
\caption{$M-\Lambda$ relation of the TOV configurations. Results are shown for the cases without DM and without the $\sigma$-$\delta$ coupling, and with the $\sigma-\delta$ coupling but without DM. For comparison, the cases with $k_F^{\text{DM}}=30, 50$~MeV, which exhibit more significant DM effects are also shown. The legend follows the same convention as in FIG.\ref{fig:MR comparison}.}
\label{fig:MT comparison}
\end{figure*}

We adopt RMF matter to describe the NS core, using the parameters1 in Table.\ref{tab:comparison} as the baseline. As illustrated in Sec.\ref{sec:bayes}, we consider representative combinations with fixed values $E_{\text{sym}}=27, 30, 33, 36$~MeV and $L=40, 50, 60, 70$~MeV. For a given values of $E_{\text{sym}}$ at $n_0$, the slope $L$ is allowed to continuously vary within the range $40\sim70$~MeV; conversely, for a fixed $L$, $E_{\text{sym}}$ is varied within $27\sim36$~MeV. Using the expressions for $E_{\text{sym}}$ and $L$ Eq.(\ref{eq:analytical Esym})(\ref{eq:Lsym no coupling}), the isovector coupling constants $C_{\rho}^2$ and $C_{\delta}^2$ are determined. The EOS for the NS core is then constructed by solving the meson field equations together with the charge neutrality and the chemical equilibrium condition. To obtain a complete NS EOS, the crust must also be included. For the inner crust, we adopt a polytropic form $P=A+B\varepsilon^{4/3}$, while for the outer crust, the BPS\cite{baym1971ground} EOS was employed.
Under the uncertainties in $(E_{\text{sym}})_{n_0}$ and $L_{n_0}$, we compute the M-R relations of NS corresponding to the maximum-mass configurations(TOV configurations), both with and without DM, and the situation for $\sigma-\delta$ coupling is also included. The corresponding results are shown in FIG.\ref{fig:MR comparison} and FIG.\ref{fig:DM MR comparison}, $M-\Lambda$ relation also shown in FIG.\ref{fig:MT comparison}

FIG.\ref{fig:MR comparison} presents the $M-R$ relations of the TOV configurations obtained with and without the $\sigma-\delta$ coupling with the same range of variations in NM properties($(E_{\text{sym}})_{n_0}$ and $L_{n_0}$). Several representative comparison points of the $M-R-\Lambda$ relations are listed in Table.\ref{tab:MRTcomparison}. For the same values of $(E_{\text{sym}})_{n_0}$ and $L_{n_0}$, M obtained without the coupling is systematically smaller than their counterparts with the coupling, whereas the corresponding R and $\Lambda$ are generally larger. This shows a nontrivial effect of the $\sigma-\delta$ coupling on the NS structure. The discussion in SEC.\ref{sec:Esym} indicates that the primary effect of the $\sigma-\delta$ coupling on $E_{\text{sym}}$ is softening. In general, a softer EOS leads to a smaller maximum mass of NS, the radius and the tidal deformability, as shown in FIG.\ref{fig:DM MR comparison} and FIG.\ref{fig:MT comparison} for the case with DM.

\begin{figure}
    \includegraphics[width=\linewidth]{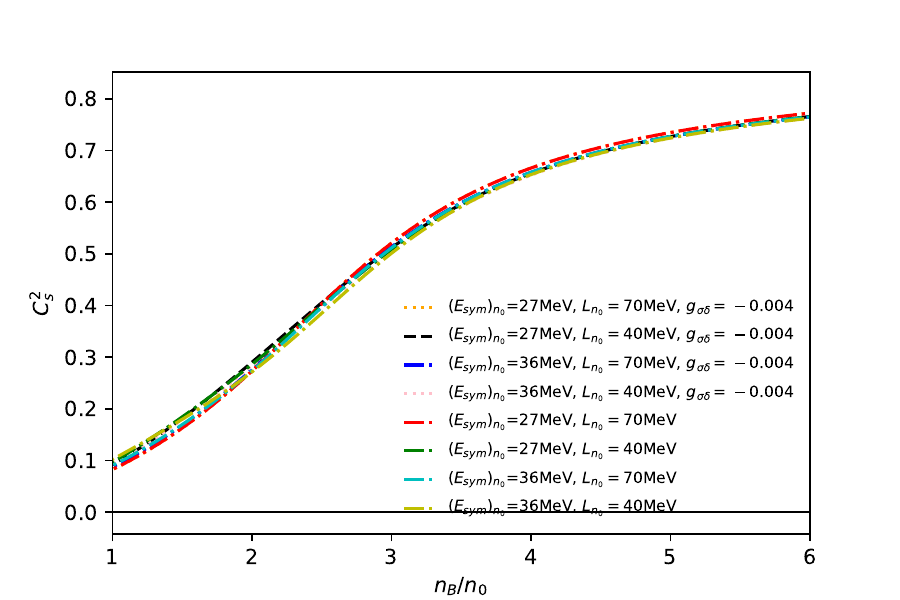}
    \caption{Speed of sound of different combinations of $(E_{\text{sym}})_{n_0}$ and $L_{n_0}$.}
    \label{fig:speedofsound}
\end{figure}

In the present case, the softening of $E_{\text{sym}}$ with the $\sigma-\delta$ coupling is mainly concentrated in the density above $n_0$, which coincides with the core-density regime for the structure of NS EOS considered in this work. Consequently, the EOS is not affected by the stiffness crossing behavior of $E_{\text{sym}}$ around $n_0$, where the stiffness ordering for different $L_{n_0}$ is reversed. From the corresponding behavior of the speed of sound in the EOS of NS, it is hardly to identify a clear trend, as shown in FIG.\ref{fig:speedofsound}.

\begin{figure}
    \includegraphics[width=\linewidth]{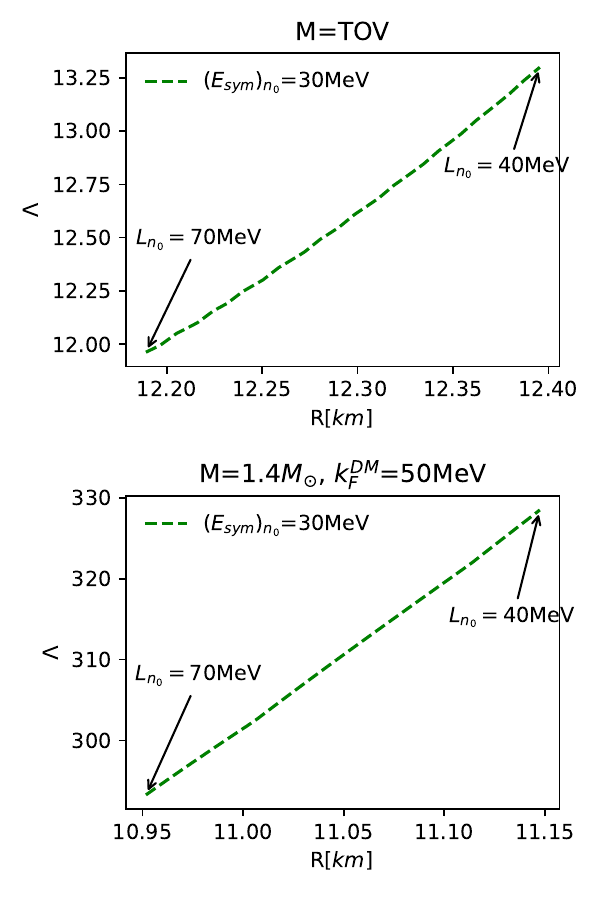}
    \caption{$R-\Lambda$ relation of TOV configurations and the canonical NS with fixed $(E_{\text{sym}})_{n_0}=30$MeV.}
    \label{fig:RT comparison}
\end{figure}

Furthermore, in the absence of DM, FIG.\ref{fig:MR comparison} and the first panel of FIG.\ref{fig:RT comparison} suggest that, for a fixed $(E_{\text{sym}})_{n_0}$, the incerasing $L_{n_0}$ generally leads to smaller mass, ridius and tidal deformability, we already know as $L_{n_0}$ increases, $E_{\text{sym}}$ becomes softer, hence the stiffness of $E_{\text{sym}}$ can be regarded as a proxy for the stiffness of the EOS. However, this trend is reversed once DM effects are introduced. FIG.\ref{fig:DM MR comparison} presents the corresponding $M-R$ relations for $k_F^{DM}=0,10,30$ and 50~MeV. Taking the blue dashed line corresponding to $(E_{\text{sym}})_{n_0}=36$~MeV as an example, the slope of the $M-R$ curve remains positive when $k_F^{\text{DM}}=0$ or sufficiently small. For $k_F^{DM}=50$~MeV, the slope of the $M-R$ curve remains negative throughout the range of $L_{n_0}$ from 40 to 70~MeV. This means that, as $E_{\text{sym}}$ becomes softer, the conventional trend of simultaneous decrease in M, R and $\Lambda$ replaced by R and $\Lambda$ decrease while M increases. Although this conclusion is obtained from the comparisons performed at fixed $k_F^{\text{DM}}$, 
the overall trend that the NS mass, the radius and the tidal deformability decrease with the deeply EOS softening remains valid as $k_F^{\text{DM}}$ increases.

As discussed above for the effects of the $\sigma-\delta$ coupling, a similar behavior is observed at $g_{\sigma\delta}=-0.004$, namely, an increase in M accompanied by reductions in both R and $\Lambda$. It is not a common phenomenon within the range of current research variations, we call it \textbf{abnormal softening}. 

The slope of MR curves exhibit a systematically decreasing trend, and further indicates that the DM effects constitute a non-negligible component in studies of the NS properties. As pointed out by Das et al.\cite{Das:2018frc}, the softening of the EOS induced by DM allows RMF NS models based on the NL3 parameter set\cite{lalazissis1997new} to satisfy the observational constraints($\Lambda_{1.4} \leqslant 580$) from GW170817\cite{170817}. Our result is the same as Das, with $k_F^{\text{DM}}=50$MeV, the tidal deformability for the canonical NS($\Lambda_{1.4}$) smaller than 580. To further examine this issue, we performed an additional set of calculations with $k_F^{\text{DM}}=40$MeV, Table.\ref{tab:1.4RTcomparison} lists the corresponding R and $\Lambda$ for $[(E_{\text{sym}})_{n_0},L_{n_0}]=[30,50]$MeV. Although $\Lambda$ exhibits moderate increases or decreases depending on the different values of $[(E_{\text{sym}})_{n_0},L_{n_0}]$, it is unchanged that the constraint from GW170817 can only be satisfied when $k_F^{\text{DM}}>40$MeV.

For the fixed $L_{n_0}$, increasing $(E_{\text{sym}})_{n_0}$ from 27 to 36~MeV, our results(the second panel in FIG.\ref{fig:EsymLcomparison1}) indicate a progressive stiffening of $E_{\text{sym}}$. Such a stiffening is generally expected to be accompanied by increases in M, R and $\Lambda$, while, as can be inferred from the left panel of FIG.\ref{fig:MR comparison} together with the trend indicated by the dotted curve in the upper-left panel of FIG.\ref{fig:MT comparison}, R and $\Lambda$ increase simultaneously, whereas M decreases, which is contrary to the outcome from the abnormal softening discussed above. We also call this behavior abnormal stiffening. The anomalous behavior of M-R-$\Lambda$ relation induced by variations in the stiffness of $E_{\text{sym}}$ will be further substantiated in the section on the curvature invariants(SEC.\ref{sec:curvature}).

In FIG.\ref{fig:RT comparison}, for the fixed $(E_{\text{sym}})_{n_0}$, both $R-\Lambda$ relation of the TOV configurations and that of the canonical NS exhibit decreasing R and $\Lambda$ with increasing $L_{n_0}$. Under the same scenario, the results of Zhu et al.\cite{zhu2018neutron}, with the increasing $L_{n_0}$, the radius of the canonical NS($R_{1.4}$) becomes larger, whereas the corresponding $\Lambda_{1.4}$ decreases anomalously, which is contrary to the expectation that a larger stellar radius generally associates with a larger $\Lambda$. Their explanation for this phenomenon is the matching between the different EOSs of the NS core and the same inner crust EOS\cite{zhu2018neutron}. 

Consistent with their conclusions, $\Lambda$ obtained in this work also decreases with increasing $L_{n_0}$. The difference is that their results exhibit an anomalous increase in R. However, the decreasing radii obtained in this work remain consistent with the expectation shown in FIG.\ref{fig:RT comparison}: The tidal deformability exhibits a positive correlation with the stellar radius. The larger values of $L_{n_0}$ correspond to the softer EOS, which naturally leads the smaller values of R and $\Lambda$.

Here, we give a possible interpretation of the anomalous results reported by Zhu et al.\cite{zhu2018neutron}: The trend of their $R-\Lambda$ relation for the canonical NS with increasing $L_{n_0}$ may also be closely related to the overall behavior of $E_{\text{sym}}$. As shown in the first panel of FIG.\ref{fig:EsymLcomparison1}, for the fixed $(E_{\text{sym}})_{n_0}$, increasing $L_{n_0}$ leads to a stiffer $E_{\text{sym}}$ with $n_B<n_0$, while $E_{\text{sym}}$ becomes softer with $n_B>n_0$. The dependence of $E_{\text{sym}}$ on $L_{n_0}$ obtained by Zhu exhibits an opposite trend compared to our results. In general, the central density of the canonical NS(without DM) is approximately $2n_0$, while the matching density between the NS core and the inner crust in their work is $0.5 n_0$, which indicates that a substantial portion of the \textbf{stiffness crossing} for $E_{\text{sym}}$ is contained within the density range relevant to the NS EOS.

In our calculations, since the density of the inner crust-core matching is fixed at $n_0$, the stiffness crossing associated with the softening-stiffening reversal of $E_{\text{sym}}$ is avoided. As a result, the resulting $R-\Lambda$ relation follows the conventional expectation. Nevertheless, the overall behavior of $E_{\text{sym}}$ in the vicinity of $n_0$ still requires a comprehensive investigation. Such an analysis is beyond the scope of this work and will be addressed in a future study.

\begin{table*}[htbp]
\caption{Some values of MR$\Lambda$ in FIG.\ref{fig:MR comparison}, FIG.\ref{fig:DM MR comparison} and FIG.\ref{fig:MT comparison}. $[E_{\mathrm{sym}},L]_{n_0}$ represents the values at $n_0$. The second and the third rows present the corresponding MR$\Lambda$ values for the cases without and with the $\sigma-\delta$ coupling, respectively. The remaining results represent the values with the admixture of DM.}
\begin{ruledtabular}
\begin{tabular}{ccccc}
 $[M(M_{\odot}),R(\mathrm{km}),\Lambda]~ \backslash ~[E_{\mathrm{sym}},L]_{n_0}$ & (27,70)MeV & (27,40)MeV & (36,70)MeV & (36,40)MeV \\
\hline
no coupling                & $(2.435, 12.118, 11.434)$ & $(2.452, 12.376, 12.844)$ & $(2.429, 12.252, 12.786)$ & $(2.432, 12.392, 13.898)$ \\
$g_{\sigma \delta}=-0.004$ & $(2.435, 12.116, 11.452)$ & $(2.454, 12.347, 12.493)$ & $(2.430, 12.233, 12.579)$ & $(2.435, 12.345, 13.327)$ \\
$k_F^{DM}=30$MeV           & $(2.340, 11.498, 10.296)$ & $(2.351, 11.705, 11.477)$ & $(2.332, 11.604, 11.433)$ & $(2.331, 11.706, 12.306)$ \\
$k_F^{DM}=50$MeV           & $(2.082, ~9.949,~~8.112)$ & $(2.084, 10.042, ~8.735)$ & $(2.072, ~9.986, ~8.740)$ & $(2.068, 10.029, ~9.215)$ \\
\end{tabular}
\end{ruledtabular}
\label{tab:MRTcomparison}
\end{table*}

\begin{table*}[htbp]
\caption{The radius and $\Lambda$ of a 1.4$M_{\odot}$ neutron star are presented with $[(E_{\text{sym}})_{n_0},L_{n_0}]=[30,50]$MeV, and additional calculations performed for the case $k_F^{DM}=40$MeV. }
\begin{ruledtabular}
\begin{tabular}{ccccccc}
   & no coupling  & $g_{\sigma \delta}=-0.004$ & $k_F^{DM}=10$MeV & $k_F^{DM}=30$MeV & $k_F^{DM}=40$MeV & $k_F^{DM}=50$MeV \\
\hline
$[R(\mathrm{km}),\Lambda]$ & $(14.537, 1683.443)$ & $(14.413, 1600.262)$ & $(14.493, 1653.622)$ & $(13.481, 1070.680)$ & $(12.374, 633.334)$ & $(11.084, 316.709)$\\
\end{tabular}
\end{ruledtabular}
\label{tab:1.4RTcomparison}
\end{table*}

\subsection{Curvature}
\label{sec:curvature}

\begin{figure}
    \includegraphics[width=\linewidth]{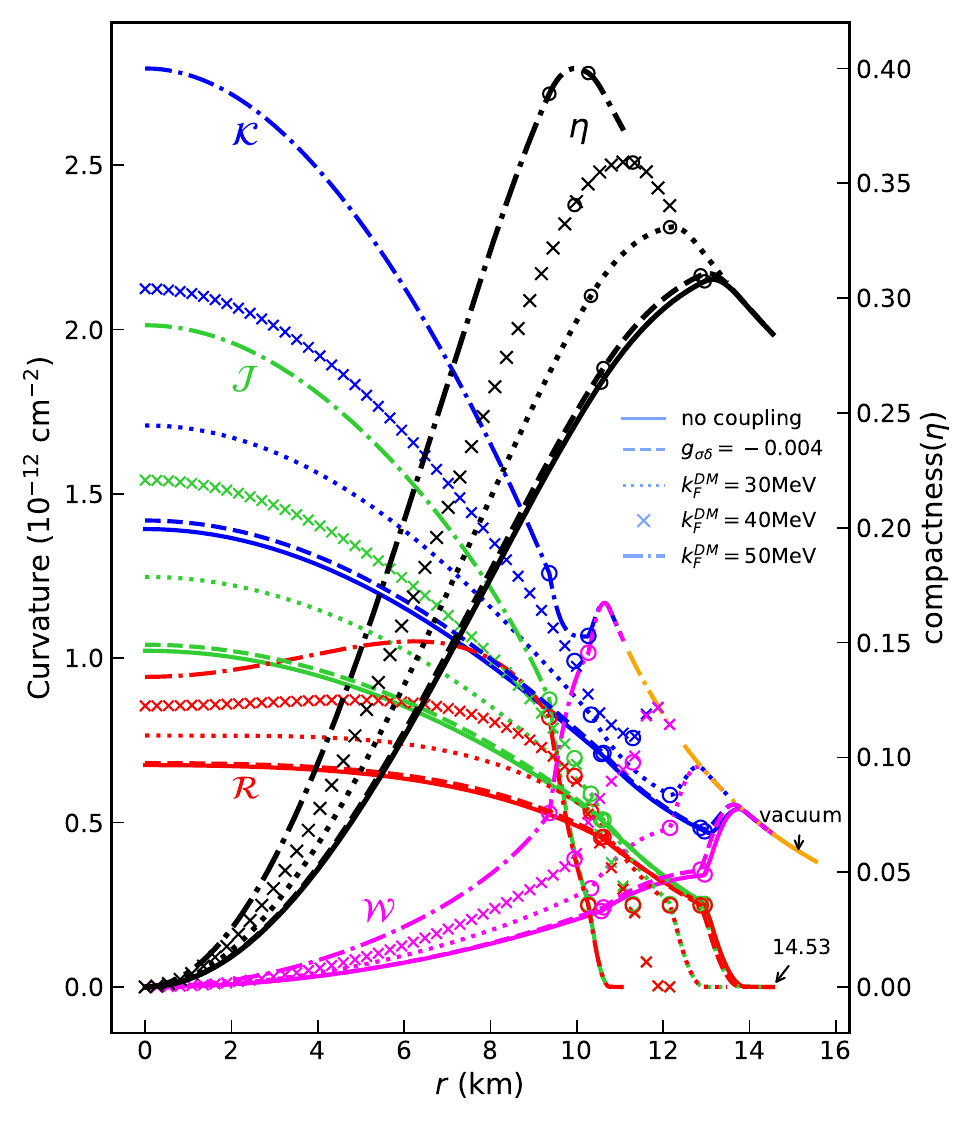}
    \caption{The radial variation of curvatures for $[(E_{\text{sym}})_{n_0}, L_{n_0}]=[30,50] \text{MeV}$, $\mathcal{K}$(blue), $\mathcal{J}$(limegreen), $\mathcal{R}$(red), $\mathcal{W}$(magenta), and the compactness $\eta$(black). The linestyle of the curve is shown in the legend. The right vertical axes corresponds to $\eta$, curvatures and compactness share the same horizontal axis r(km).}
    \label{fig:curvature1}
\end{figure}
We compute the radial profiles of various curvature invariants for a canonical NS with $[(E_{\text{sym}})_{n_0}, L_{n_0}]=[30,50]$~MeV with different configurations($\sigma-\delta$ coupling, admixture of DM with different $k_F^{\text{DM}}$), the results are shown in FIG.\ref{fig:curvature1}. The results are presented for three cases: (i)~without $\sigma-\delta$ coupling, (ii)~with $g_{\sigma\delta}=-0.004$, and (iii)~with DM included but without the $\sigma-\delta$ coupling. The two open circles marked on each curve denote the interfaces between the NS core and the inner crust, and between the inner and outer crust, respectively. For the Kretschmann scalar $\mathcal{K}$ and the full contraction of Weyl tensor $\mathcal{W}$, the orange segments beyond the stellar surface represent the extension of the calculations into the vacuum up to 1km. 14.53km represents the radius of a canonical NS without the $\sigma-\delta$ coupling. 

For all cases considered, both $\mathcal{K}$ and the full contraction of Ricci tensor $\mathcal{J}$ have the maximum values at the stellar center. Consistent with the results of Das et al.\cite{das2021impacts}, the Ricci scalar $\mathcal{R}$ also reaches its maximum at the core for $k_F^{DM}\leq 40$MeV, while $k_F^{DM}> 40$MeV, $\mathcal{R}$ reaches its maximum at a radius of approximately 6$\sim$7~km, and subsequently decreases to zero at the stellar surface. According to the expression for $\mathcal{R}$ Eq.(\ref{eq:Ricci scalar}), in the core, both the pressure and the energy density are extremely large, however, an excessively soft EOS implies that a high energy density is accompanied by a relatively small pressure, hence the pressure inside the NS exhibits a much steeper variation from the core (particularly near the inner crust) to the crust and finally to the stellar surface. At the same time, a larger compactness associated with a softer EOS leads to a smaller star radius, further amplifying the separation between the energy density and the pressure contributions. 
As a result, the value of $\mathcal{R}$ at high energy densities can be smaller than that in regions closer to the crust. In other words, for a more compact NS, the pressure drops more rapidly than the energy density at radii of approximately 6$\sim$7~km. This difference in their radial evolution causes $\mathcal{R}$ to increase initially and then decrease toward zero as the stellar surface approached.

$\mathcal{R}$ and $\mathcal{J}$, which are clearly separated in the core for all cases, begin to converge in the inner crust and become nearly indistinguishable in the outer crust. Combined with Eqs.(\ref{eq:Ricci scalar}) and (\ref{eq:Ricci tensor}), this behavior reflects the transition from pressure-dominated in the core to energy-density-dominated in the crust. Physically, it originates from the rapid decrease in pressure inside the NS. Outside the NS, where the pressure and the energy density vanish, $\mathcal{K}$ and $\mathcal{W}$ become identical. $\mathcal{W}$ increases gradually from the core, reaches its maximum within the outer crust, and attains its vacuum value at the stellar surface, $\frac{4 \sqrt{3} G M}{r^3 c^2}$, subsequently decreases in the exterior vacuum region. According to the interpretation of Das \cite{das2021impacts}, this behavior originates from the energy density approaches a diffuse-state near the crustal region, causing $\mathcal{W}$ attain its maximum in the outer crust. Moreover, this effect becomes more pronounced for NS with higher compactness.

\begin{figure*}[t]
    \centering
    \includegraphics[width=0.48\linewidth]{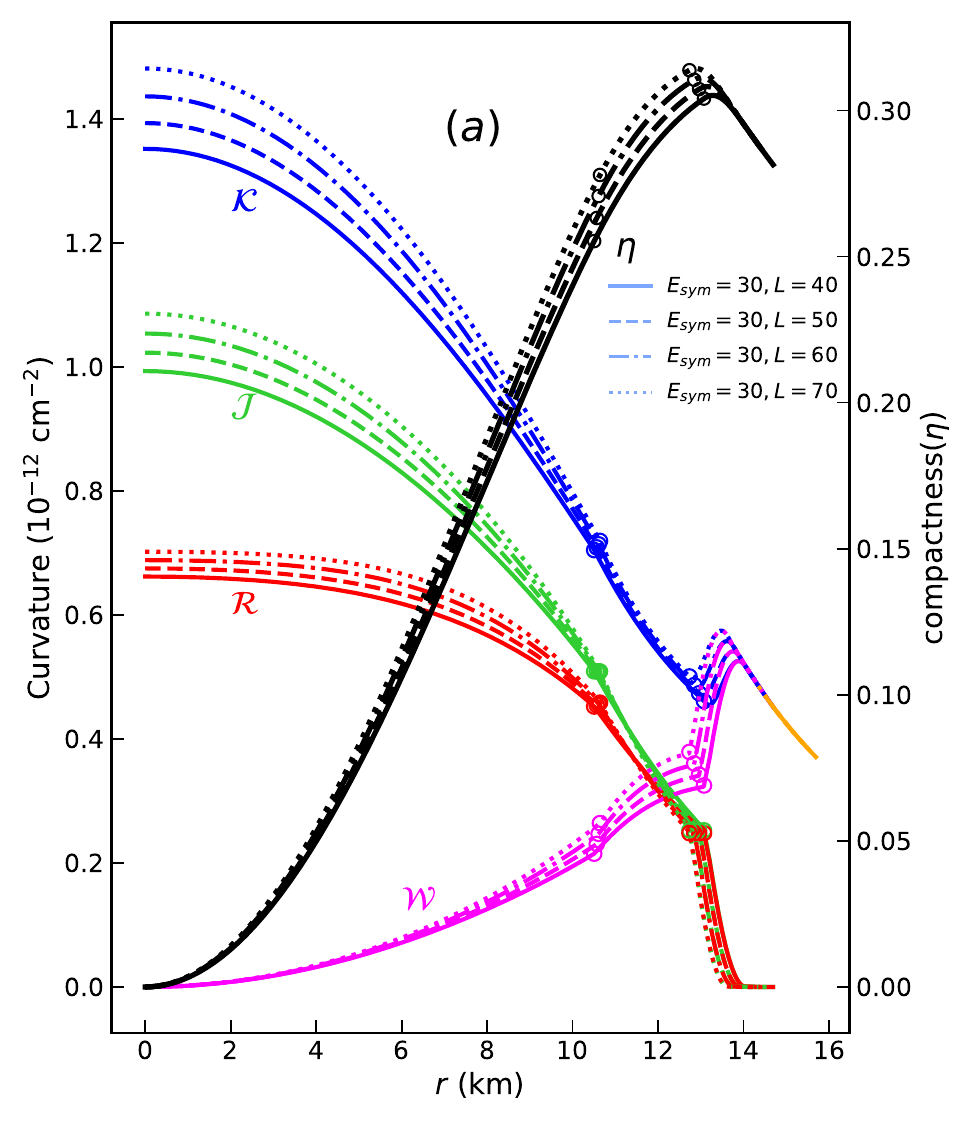}
    \hfill
    \includegraphics[width=0.48\linewidth]{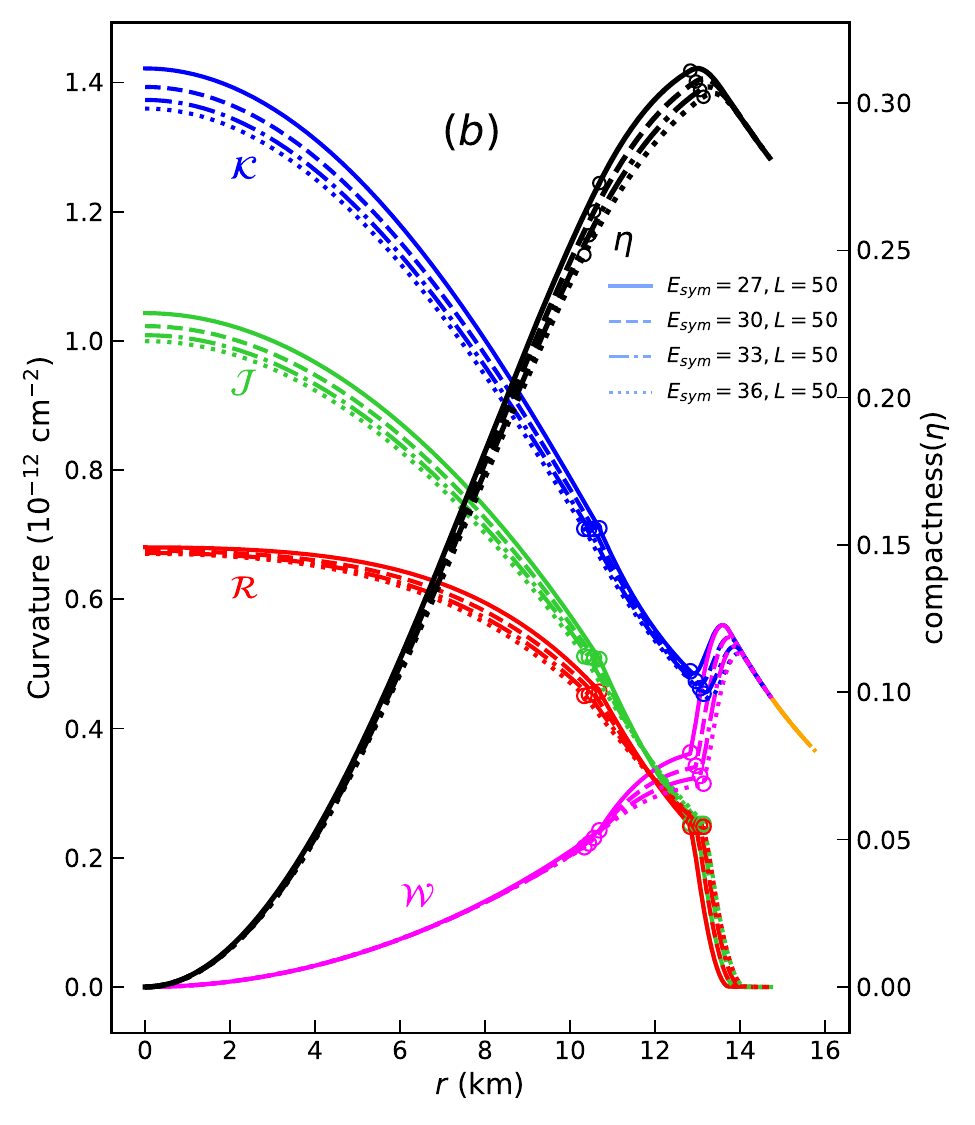}
    \caption{Different combinations of $(E_{\text{sym}})_{n_0}$ and $L_{n_0}$.
        $(a)$: Curvature with different $L_{n_0}$ at fixed $(E_{\text{sym}})_{n_0}=30$MeV.
        ~$(b)$: Curvature with different $(E_{\text{sym}})_{n_0}$ at fixed $L_{n_0}=50$MeV.
    The right vertical axes correspond to compactness($\eta$).}
    \label{fig:curvature2}
\end{figure*}

In FIG.\ref{fig:curvature2}, we further calculate the curvature quantities and the compactness of the canonical NS corresponding to different combinations of $(E_{\text{sym}})_{n_0}$ and $L_{n_0}$. From the previous analysis, it is known that for the symmetry-energy $E_{\text{sym}}(n_B)$, obtained from the Lagrangian Eq.(\ref{eq:lagrangian}), at the fixed $\left(E_{\text {sym}}\right)_{n_0}$, increasing $L_{n_0}$ softens $E_{\text{sym}}\left(n_B\right)$ above $n_0$. Similarly, at the fixed $L_{n_0}$, decreasing $\left(E_{\text {sym}}\right)_{n_0}$ also results in a softer $E_{\text{sym}}(n_B)$.

The corresponding curvature profiles shown in FIG.\ref{fig:curvature2} indicate that a softer $E_{\text{sym}}(n_B)$ generally leads to larger curvature values, while a stiffer $E_{\text{sym}}(n_B)$ corresponds to a smaller curvature values. We underscore again that the relation between the stiffness of $E_{\text{sym}}(n_B)$ and the saturation properties $(E_{\text{sym}})_{n_0}$ and $L_{n_0}$ is rather complex. Based on the QMC\cite{guichon1988possible, Saito:2005rv} and QMF\cite{toki1998quark, shen2000quark, shen2002study} models within the RMF framework, Zhu et al.\cite{zhu2018neutron} obtained an opposite trend in which $E_{\text{sym}}(n_B)$ becomes stiffer above $n_0$ and softer below $n_0$ with increasing $L_{n_0}$.

It can therefore be inferred that, if their EOS were used in the curvature calculations, larger values of curvature quantities would yield conclusions regarding the saturation properties of nuclear matter that are contrary to those obtained in the present work.

In addition, we note that regardless of the degree of EOS softening or how large $\mathcal{K}$ and $\mathcal{W}$ become at the stellar surface, both curvature outside the stellar are described by the same vacuum expression at any given radius r, $\frac{4 \sqrt{3} G M}{r^3 c^2}$. Since the canonical neutron star possesses the same gravitational mass(1.4$M_{\odot}$), the exterior curvature profiles corresponding to different values of $k_F^{\text{DM}}$ (i.e., different degrees of EOS softening) evolve along the same trajectory outside the stellar. This conclusion remains unchanged regardless of the stiffness of the EOS or the nuclear model adopted.

For each distinct stellar mass, there exists an independent exterior evolution curve, and the same argument also applies to the evolution of the compactness. Once the stellar mass fixed, the surface curvature depends only on the stellar radius. This also explains the enormous difference between the surface curvature of neutron stars and that of the Sun, $\mathcal{K}(R) / \mathcal{K}_{\odot} \approx 10^{14}$.

Furthermore, the relationship established above between the stiffness of the symmetry energy and the magnitude of the curvature quantities provides additional support for the anomalous behavior discussed in SEC.\ref{sec:MRLambda relation}. In FIG.\ref{fig:curvature1}, the solid curve corresponding to the case without the $\sigma-\delta$ coupling and without DM is taken as the reference. Compared with this baseline, the cases including DM($k_F^{\text{DM}}=30,40,50$MeV) but excluding the $\sigma-\delta$ coupling exhibit larger values of curvature quantities and compactness, indicating a softening of the EOS. On the other hand, the dashed line, which represents the case with only the $\sigma-\delta$ coupling included, lies above the reference curve. According to the curvature$-$$E_{\text{sym}}$ correlation, this result implies the increase in stellar mass accompanied by decreases in radius and tidal deformability discussed in SEC.\ref{sec:MRLambda relation} corresponds to a softening of the EOS. In the present framework, this behavior can be further interpreted as a consequence of the softening of $E_{\text{sym}}$.

Also, for different combinations of $(E_{\text{sym}})_{n_0}$ and $L_{n_0}$, FIG.\ref{fig:curvature2} further shows the sensitivity of the curvature quantities to the degree of the stiffness of $E_{\text{sym}}$. In particular, the increase in $(E_{\text{sym}})_{n_0}$ at fixed $L_{n_0}$, which was discussed in SEC.\ref{sec:MRLambda relation} to be associated with a decrease in the stellar mass and simultaneous increases in the radius and the tidal deformability, corresponds to an abnormal stiffening. As shown in FIG.\ref{fig:curvature2}(b), the curves associated with larger values of $(E_{\text{sym}})_{n_0}$ lie systematically below those corresponding to the smallest $(E_{\text{sym}})_{n_0}$ in terms of both the curvature quantities and the compactness. According to the lower panel of FIG.\ref{fig:EsymLcomparison1}, larger $(E_{\text{sym}})_{n_0}$ corresponds to a stiffer $E_{\text{sym}}$, Therefore, the curvature analysis independently confirms that the decrease in mass accompanied by increases in radius and tidal deformability discussed in SEC.\ref{sec:MRLambda relation} indeed corresponds to a stiffening of $E_{\text{sym}}$. 
This provides a self-consistent justification for using the stiffness of the symmetry energy as a proxy for the stiffness of the EOS, especially in situations where the sound-speed behavior alone is insufficient to distinguish between the different nuclear matter properties for EOSs.

\section{SUMMARY}
\label{sec:summary}

In the era of multimessenger astronomy, combined with terrestrial nuclear experiments, the unknown equation of state (EOS) of supranuclear matter is expected to be more precisely constrained. In this work, within the framework of the four-meson($\sigma$, $\omega$, $\boldsymbol{\rho}$, $\boldsymbol{\delta}$) model, we employ Bayesian inference to reproduce nuclear matter properties associated with isoscalar parameters, effectively reducing the incompressibility. Specifically, a nonlinear $\sigma$-$\delta$ coupling with a coupling constant $g_{\sigma\delta} = -0.004$ is introduced into the Lagrangian Eq.(\ref{eq:lagrangian}). Analytical density-dependent expressions for the symmetry energy $E_{\mathrm{sym}}(n_B)$, its slope $L(n_B)$, and the incompressibility at saturation density \( K_0 \) are derived. Using the experimental/empirical constraints on $E_{\mathrm{sym}}(n_0)$ and $L(n_0)$, we determine the allowed distributions of the isovector parameters and propagate their uncertainties into the nuclear matter EOS. Additionally, considering the possible effects of dark matter (DM), we investigate the evolution of the neutron-star mass–radius (M–R) relation, tidal deformability, and curvature invariants under these parametric uncertainties.

Our results reveal a nontrivial feature. A direct comparison between the M-R relation with the $\sigma$-$\delta$ coupling and those including DM effects—which typically soften the EOS—suggests that the $\sigma$-$\delta$ coupling stiffens the EOS. However, a combined analysis of the M-R and M-$\Lambda$ relations demonstrates that the $\sigma$-$\delta$ coupling actually softens the EOS. This softening is anomalous: it reduces both the tidal deformability and the stellar radius, while simultaneously increasing the stellar mass. For example, at a fixed $ E_{\mathrm{sym}}(n_0) = 36 \text{MeV}$ and $L_{n_0} = 40 \, \text{MeV}$, the maximum neutron star mass increases by approximately 0.003 \( M_\odot \) when the $\sigma$-$\delta$ coupling is included, whereas the radius decreases by about 0.05 km. A similar anomalous behavior is observed in the M–R relation for $k_F^{\mathrm{DM}} = 50 \, \text{MeV}$ with $E_{\mathrm{sym}}(n_0) = 36 \, \text{MeV}$. For smaller values of $k_F^{\mathrm{DM}}$, the EOS softening induced by increasing \( L_{n_0} \) leads to simultaneous reductions in both mass and radius (e.g., a decrease of 0.017 \( M_\odot \) and 0.261 km for \( k_F^{\mathrm{DM}} = 10 \, \text{MeV}, E_{\text{sym}}(n_0)=27~\text{MeV} \) ). In contrast, for \( k_{\mathrm{DM}}^F = 50 \, \text{MeV} \), the mass increases while the radius decreases. We emphasize that this abnormal softening originates from the additional softening of \( E_{\mathrm{sym}}(n_B) \) above \( n_0 \), driven by the continuous increase of \( L_{n_0} \).

Concerning the tidal deformability of the canonical neutron star, our conclusions align with those of Das et al.\cite{Das:2018frc}. In their study, using the NL3 parameter set\cite{lalazissis1997new} with DM effects, the GW170817 constraint was satisfied for \( k_F^{\mathrm{DM}} = 50 \, \text{MeV} \). Our results further show that for \( k_F^{\mathrm{DM}}= 50 \, \text{MeV} \), the dimensionless tidal deformability \( \Lambda_{1.4} \) falls within the range \( 280\sim 340\) with current variation range in nuclear matter uncertainties($E_{\text{sym}}(n_0) \in [27,36]~\text{MeV}, L_{n_0} \in [40,70]~ \text{MeV}$) , consistent with the GW170817 bounds.

Additionally, we propose a possible explanation for the anomalous behavior reported by Zhu et al.\cite{zhu2018neutron}, where increasing \( L_{n_0} \) leads to larger radii but smaller tidal deformability for the canonical neutron star, while in our result, an increase in \( L_{n_0} \) from 40 MeV to 70 MeV at fixed values of \( E_{\mathrm{sym}}(n_0) \) results in a $\sim$0.19 km decrease in radius and a $\sim$35-unit decrease in \( \Lambda_{1.4} \)(\( k_F^{\mathrm{DM}}=50 \, \text{MeV} \) ). This phenomenon may be closely related to the density matching region adopted in the NS EOS construction. The behavior within this interval is likely associated with the overall behavior of \( E_{\mathrm{sym}}(n_B) \) around \( n_0 \). In our EOS matching procedure, we avoid the crossing behavior of \( E_{\mathrm{sym}}(n_B) \) around $n_0$ for different values of $L_{n_0}$, whereas such crossing behavior is not excluded in the calculations of Zhu et al., Therefore, the difference in the overall impact of the crossing behavior of \( E_{\mathrm{sym}}(n_B) \) may explain the anomalous relationship between the NS radius and tidal deformability.

Moreover, through comparisons with the results of Ref.\cite{zhu2018neutron} and Ref.\cite{Li:2022okx}, we find that the influence of varying \( L_{n_0} \) at fixed \( E_{\mathrm{sym}}(n_0) \) on the behavior of \( E_{\mathrm{sym}}(n_B) \) above and below \( n_0 \) depends sensitively on the isoscalar parameter set employed. For example, the behavior of \( E_{\mathrm{sym}}(n_B) \) around \( n_0 \) in this work is opposite to that reported by Zhu et al.\cite{zhu2018neutron}. A more detailed investigation of this issue will be carried out in future work.

For the curvature invariants, our results indicate that the effects of DM, the $\sigma$-$\delta$ coupling, and the different combinations of \( E_{\mathrm{sym}}(n_0) \) and \( L_{n_0} \) mainly influence the curvature indirectly through the variation of the \( E_{\mathrm{sym}}(n_B) \) stiffness. The relation between \( E_{\mathrm{sym}}(n_B) \) and the curvature invariants demonstrates that the stiffness of \( E_{\mathrm{sym}}(n_B) \) can effectively characterize the stiffness of the neutron-star EOS. This provides a complementary diagnostic tool to the sound-speed analysis, which may be insufficient to distinguish different EOS behaviors in certain cases. Considering the expressions of the curvature scalars $\mathcal{K}$ and $\mathcal{W}$ in vacuum, EOSs with different stiffnesses but the same stellar mass share a common curvature evolution trajectory outside the star. Similar conclusions also apply to the compactness (or equivalently, the gravitational redshift). Once the mass is known, precise measurements of the surface curvature would provide valuable constraints on the neutron-star radius and help reduce the associated uncertainties. Although the behavior of \( E_{\mathrm{sym}}(n_B) \) around \( n_0 \) may depend on the specific model or parameter set employed, variations in \( L_{n_0} \) generally produce a more noticeable effect on the NS radius compared with changes in \( E_{\mathrm{sym}}(n_0) \). This may provide an important avenue for a deeper understanding of dense nuclear matter in future studies.

\begin{acknowledgments}
We are grateful to Cheng-Jun Xia, Lie-wen Chen, Jin-Niu Hu, Sophia Han, and Ang Li for the valuable and insightful discussions. This work is supported in part by the National Key Research and Development Program of China under Contract No.2022YFA1604900. This work is also partly supported by the National Natural Science Foundation of China(NSFC) under Grants No.12435009, and No.12275104.
\end{acknowledgments}

\bibliographystyle{apsrev4-2}
\bibliography{bibliography}

@article{tolman1939static,
  title={Static solutions of Einstein's field equations for spheres of fluid},
  author={Tolman, Richard C},
  doi={https://doi.org/10.1103/PhysRev.55.364},
  journal={Physical Review},
  volume={55},
  number={4},
  pages={364},
  year={1939},
  publisher={APS}
}

@article{oppenheimer1939massive,
  title={On massive neutron cores},
  author={Oppenheimer, J Robert and Volkoff, George M},
  doi={https://doi.org/10.1103/PhysRev.55.374},
  journal={Physical Review},
  volume={55},
  number={4},
  pages={374},
  year={1939},
  publisher={APS}
}

@article{gambhir1990relativistic,
  title={Relativistic mean field theory for finite nuclei},
  author={Gambhir, YK and Ring, P and Thimet, Ann},
  doi={https://doi.org/10.1016/0003-4916(90)90330-Q},
  journal={Annals of Physics},
  volume={198},
  number={1},
  pages={132--179},
  year={1990},
  publisher={Elsevier}
}

@article{ring1996relativistic,
  title={Relativistic mean field theory in finite nuclei},
  doi={10.1016/0146-6410(96)00054-3},
  author={Ring, Peter},
  journal={Progress in Particle and Nuclear Physics},
  volume={37},
  pages={193--263},
  year={1996},
  publisher={Elsevier}
}

@article{dutra2014relativistic,
  title={Relativistic mean-field hadronic models under nuclear matter constraints},
  author={Dutra, M and Louren{\c{c}}o, O and Avancini, SS and Carlson, BV and Delfino, A and Menezes, DP and Provid{\^e}ncia, C and Typel, S and Stone, JR},
  doi={https://doi.org/10.1103/PhysRevC.90.055203},
  journal={Physical Review C},
  volume={90},
  number={5},
  pages={055203},
  year={2014},
  publisher={APS}
}

@article{johnson1955classical,
  title={Classical field theory of nuclear forces},
  author={Johnson, MH and Teller, Edward},
  doi= { https://doi.org/10.1103/PhysRev.98.783},
  journal={Physical Review},
  volume={98},
  number={3},
  pages={783},
  year={1955},
  publisher={APS}
}

@article{durr1956relativistic,
  title={Relativistic effects in nuclear forces},
  author={D{\"u}rr, Hans-Peter},
  doi = {https://doi.org/10.1103/PhysRev.103.469},
  journal={Physical Review},
  volume={103},
  number={2},
  pages={469},
  year={1956},
  publisher={APS}
}

@article{walecka1974theory,
  title={A theory of highly condensed matter},
  author={Walecka, John Dirk},
  doi={https://doi.org/10.1016/0003-4916(74)90208-5},
  journal={Annals of Physics},
  volume={83},
  number={2},
  pages={491--529},
  year={1974},
  publisher={Elsevier}
}

@article{li2013constraining,
  title={Constraining the neutron--proton effective mass splitting using empirical constraints on the density dependence of nuclear symmetry energy around normal density},
  author={Li, Bao-An and Han, Xiao},
  doi={10.1016/j.physletb.2013.10.006},
  journal={Physics Letters B},
  volume={727},
  number={1-3},
  pages={276--281},
  year={2013},
  publisher={Elsevier}
}

@article{wang2025extended,
  title={Extended momentum-dependent interaction for transport models and neutron stars},
  author={Wang, Si-Pei and Chen, Lie-Wen},
  doi={https://doi.org/10.1103/v21s-d9dt},
  journal={Physical Review C},
  volume={112},
  number={4},
  pages={044612},
  year={2025},
  publisher={APS}
}

@article{brown2000neutron,
  title={Neutron radii in nuclei and the neutron equation of state},
  author={Brown, B Alex},
  doi={https://doi.org/10.1103/PhysRevLett.85.5296},
  journal={Physical review letters},
  volume={85},
  number={25},
  pages={5296},
  year={2000},
  publisher={APS}
}

@article{chen2005nuclear,
  title={Nuclear matter symmetry energy and the neutron skin thickness of heavy nuclei},
  author={Chen, Lie-Wen and Ko, Che Ming and Li, Bao-An},
  doi={https://doi.org/10.1103/PhysRevC.72.064309},
  journal={Physical Review C—Nuclear Physics},
  volume={72},
  number={6},
  pages={064309},
  year={2005},
  publisher={APS}
}

@article{centelles2009nuclear,
  title={Nuclear symmetry energy probed by neutron skin thickness of nuclei},
  author={Centelles, M and Roca-Maza, X and Vinas, X and Warda, M},
  doi={https://doi.org/10.1103/PhysRevLett.102.122502},
  journal={Physical review letters},
  volume={102},
  number={12},
  pages={122502},
  year={2009},
  publisher={APS}
}

@article{typel2001neutron,
  title={Neutron radii and the neutron equation of state in relativistic models},
  author={Typel, S and Brown, B Alex},
  doi={https://doi.org/10.1103/PhysRevC.64.027302},
  journal={Physical Review C},
  volume={64},
  number={2},
  pages={027302},
  year={2001},
  publisher={APS}
}

@article{Pb208,
  title={Accurate determination of the neutron skin thickness of Pb 208 through parity-violation in electron scattering},
  author={Adhikari, D and Albataineh, H and Androic, D and Aniol, K and Armstrong, DS and Averett, T and Ayerbe Gayoso, C and Barcus, S and Bellini, V and Beminiwattha, RS and others},
  doi={https://doi.org/10.1103/PhysRevLett.126.172502},
  journal={Physical review letters},
  volume={126},
  number={17},
  pages={172502},
  year={2021},
  publisher={APS}
}

@article{adhikari2022precision,
  title={Precision determination of the neutral weak form factor of Ca 48},
  author={Adhikari, D and Albataineh, H and Androic, D and Aniol, KA and Armstrong, DS and Averett, T and Ayerbe Gayoso, C and Barcus, SK and Bellini, V and Beminiwattha, RS and others},
  doi={https://doi.org/10.1103/PhysRevLett.129.042501},
  journal={Physical review letters},
  volume={129},
  number={4},
  pages={042501},
  year={2022},
  publisher={APS}
}

@article{Li:2022okx,
    author = "Li, Fan and Cai, Bao-Jun and Zhou, Ying and Jiang, Wei-Zhou and Chen, Lie-Wen",
    title = "{Effects of Isoscalar- and Isovector-scalar Meson Mixing on Neutron Star Structure}",
    eprint = "2202.08705",
    archivePrefix = "arXiv",
    primaryClass = "nucl-th",
    doi = "10.3847/1538-4357/ac5e2a",
    journal = "Astrophys. J.",
    volume = "929",
    number = "2",
    pages = "183",
    year = "2022"
}

@article{FSUGold,
  title={Neutron-Rich Nuclei and Neutron Stars: A New Accurately Calibrated Interaction<? format?> for the Study of Neutron-Rich Matter},
  author={Todd-Rutel, BG and Piekarewicz, J},
  doi={10.1103/PhysRevLett.95.122501},
  journal={Physical review letters},
  volume={95},
  number={12},
  pages={122501},
  year={2005},
  publisher={APS}
}

@article{reed2021implications ,
  title={Implications of PREX-2 on the equation of state of neutron-rich matter},
  author={Reed, Brendan T and Fattoyev, Farrukh J and Horowitz, Charles J and Piekarewicz, Jorge},
  doi={https://doi.org/10.1103/PhysRevLett.126.172503},
  journal={Physical Review Letters},
  volume={126},
  number={17},
  pages={172503},
  year={2021},
  publisher={APS}
}

@article{boguta1977relativistic,
  title={Relativistic calculation of nuclear matter and the nuclear surface},
  author={Boguta, J and Bodmer, AR},
  doi={10.1016/0375-9474(77)90626-1},
  journal={Nuclear Physics A},
  volume={292},
  number={3},
  pages={413--428},
  year={1977},
  publisher={Elsevier}
}

@article{kubis1997nuclear,
  title={Nuclear matter in relativistic mean field theory with isovector scalar meson},
  author={Kubis, S and Kutschera, Marek},
  doi={10.1016/S0370-2693(97)00306-7},
  journal={Physics Letters B},
  volume={399},
  number={3-4},
  pages={191--195},
  year={1997},
  publisher={Elsevier}
}

@article{scognamiglio2026ultra,
  title={An ultra-high-resolution map of (dark) matter},
  doi = {https://doi.org/10.1038/s41550-025-02763-9},
  author={Scognamiglio, Diana and Leroy, Gavin and Harvey, David and Massey, Richard and Rhodes, Jason and Akins, Hollis B and Brinch, Malte and Berman, Edward and Casey, Caitlin M and Drakos, Nicole E and others},
  journal={Nature Astronomy},
  pages={1--10},
  year={2026},
  publisher={Nature Publishing Group}
}

@article{joglekar2020relativistic,
  title={Relativistic capture of dark matter by electrons in neutron stars},
  author={Joglekar, Aniket and Raj, Nirmal and Tanedo, Philip and Yu, Hai-Bo},
  doi={10.1016/j.physletb.2020.135767},
  journal={Physics Letters B},
  volume={809},
  pages={135767},
  year={2020},
  publisher={Elsevier}
}

@article{kouvaris2011constraining,
  title={Constraining Asymmetric Dark Matter through observations of compact stars},
  author={Kouvaris, Chris and Tinyakov, Peter},
  doi = {https://doi.org/10.1103/PhysRevD.83.083512},
  journal={Physical Review D—Particles, Fields, Gravitation, and Cosmology},
  volume={83},
  number={8},
  pages={083512},
  year={2011},
  publisher={APS}
}

@article{baryakhtar2017dark,
  title={Dark kinetic heating of neutron stars and an infrared window on WIMPs, SIMPs, and pure Higgsinos},
  author={Baryakhtar, Masha and Bramante, Joseph and Li, Shirley Weishi and Linden, Tim and Raj, Nirmal},
  journal={Physical review letters},
  volume={119},
  number={13},
  pages={131801},
  year={2017},
  publisher={APS}
}

@article{Bhat20,
  title={Cooling of dark-matter admixed neutron stars with density-dependent equation of state},
  author={Bhat, Sajad A and Paul, Avik},
  doi={https://doi.org/10.1140/epjc/s10052-020-8072-x},
  journal={The European Physical Journal C},
  volume={80},
  number={6},
  pages={544},
  year={2020},
  publisher={Springer}
}

@article{Quddus,
  title={GW170817 constraints on the properties of a neutron star in the presence of WIMP dark matter},
  author={Quddus, Abdul and Panotopoulos, Grigorios and Kumar, Bharat and Ahmad, Shakeb and Patra, SK},
  doi={10.1088/1361-6471/ab9d36},
  journal={Journal of Physics G: Nuclear and Particle Physics},
  volume={47},
  number={9},
  pages={095202},
  year={2020},
  publisher={IOP Publishing}
}

@article{Ivanytskyi20,
  title={Neutron stars: New constraints on asymmetric dark matter},
  author={Ivanytskyi, O and Sagun, V and Lopes, I},
  doi={ https://doi.org/10.1103/PhysRevD.102.063028},
  journal={Physical Review D},
  volume={102},
  number={6},
  pages={063028},
  year={2020},
  publisher={APS}
}

@article{Ellis18,
  title={Dark matter effects on neutron star properties},
  author={Ellis, John and H{\"u}tsi, Gert and Kannike, Kristjan and Marzola, Luca and Raidal, Martti and Vaskonen, Ville},
  doi={https://doi.org/10.1103/PhysRevD.97.123007},
  journal={Physical Review D},
  volume={97},
  number={12},
  pages={123007},
  year={2018},
  publisher={APS}
}

@article{panotopoulos2017dark,
  title={Dark matter effect on realistic equation of state in neutron stars},
  author={Panotopoulos, Grigorios and Lopes, Il{\'\i}dio},
  doi={https://doi.org/10.1103/PhysRevD.96.083004},
  journal={Physical Review D},
  volume={96},
  number={8},
  pages={083004},
  year={2017},
  publisher={APS}
}

@article{li2012too,
  title={Too massive neutron stars: The role of dark matter?},
  author={Li, Ang and Huang, Feng and Xu, Ren-Xin},
  doi={10.1016/j.astropartphys.2012.07.006},
  journal={Astroparticle Physics},
  volume={37},
  pages={70--74},
  year={2012},
  publisher={Elsevier}
}

@article{leung2011dark,
  title={Dark-matter admixed neutron stars},
  author={Leung, S-C and Chu, M-C and Lin, L-M},
  doi={ https://doi.org/10.1103/PhysRevD.84.107301},
  journal={Physical Review D—Particles, Fields, Gravitation, and Cosmology},
  volume={84},
  number={10},
  pages={107301},
  year={2011},
  publisher={APS}
}

@article{kouvaris2010can,
  title={Can neutron stars constrain dark matter?},
  author={Kouvaris, Chris and Tinyakov, Peter},
  doi={https://doi.org/10.1103/PhysRevD.82.063531},
  journal={Physical Review D—Particles, Fields, Gravitation, and Cosmology},
  volume={82},
  number={6},
  pages={063531},
  year={2010},
  publisher={APS}
}

@article{sandin2009effects,
  title={Effects of mirror dark matter on neutron stars},
  author={Sandin, Fredrik and Ciarcelluti, Paolo},
  doi={10.1016/j.astropartphys.2009.09.005},
  journal={Astroparticle Physics},
  volume={32},
  number={5},
  pages={278--284},
  year={2009},
  publisher={Elsevier}
}

@article{ciarcelluti2011have,
  title={Have neutron stars a dark matter core?},
  author={Ciarcelluti, Paolo and Sandin, Fredrik},
  doi={10.1016/j.physletb.2010.11.021},
  journal={Physics Letters B},
  volume={695},
  number={1-4},
  pages={19--21},
  year={2011},
  publisher={Elsevier}
}

@article{de2010neutron,
  title={Neutron stars as dark matter probes},
  author={de Lavallaz, Arnaud and Fairbairn, Malcolm},
  doi={https://doi.org/10.1103/PhysRevD.81.123521},
  journal={Physical Review D—Particles, Fields, Gravitation, and Cosmology},
  volume={81},
  number={12},
  pages={123521},
  year={2010},
  publisher={APS}
}

@article{dedeo2003towards,
  title={Towards new tests of strong-field gravity with measurements of surface atomic line redshifts from neutron stars},
  author={DeDeo, Simon and Psaltis, Dimitrios},
  doi={https://doi.org/10.1103/PhysRevLett.90.141101},
  journal={Physical review letters},
  volume={90},
  number={14},
  pages={141101},
  year={2003},
  publisher={APS}
}

@article{ekcsi2014does,
  title={What does a measurement of mass and/or radius of a neutron star constrain: Equation of state or gravity?},
  author={Ek{\c{s}}i, Kaz{\i}m Yavuz and G{\"u}ng{\"o}r, Can and T{\"u}rko{\u{g}}lu, Murat Metehan},
  doi = {https://doi.org/10.1103/PhysRevD.89.063003},
  journal={Physical Review D},
  volume={89},
  number={6},
  pages={063003},
  year={2014},
  publisher={APS}
}

@article{rosi2015measurement,
  title={Measurement of the gravity-field curvature by atom interferometry},
  author={Rosi, G and Cacciapuoti, L and Sorrentino, F and Menchetti, M and Prevedelli, Marco and Tino, GM},
  doi = {https://doi.org/10.1103/PhysRevLett.114.013001},
  journal={Physical Review Letters},
  volume={114},
  number={1},
  pages={013001},
  year={2015},
  publisher={APS}
}

@article{hugenholtz1958theorem,
  title={A theorem on the single particle energy in a Fermi gas with interaction},
  author={Hugenholtz, Nicolaas M and Van Hove, L{\'e}on},
  doi = {https://doi.org/10.1016/S0031-8914(58)95281-9},
  journal={Physica},
  volume={24},
  number={1-5},
  pages={363--376},
  year={1958},
  publisher={Elsevier}
}

@article{oertel2017equations,
  title={Equations of state for supernovae and compact stars},
  author={Oertel, Micaela and Hempel, Matthias and Kl{\"a}hn, Thomas and Typel, Stefan},
  doi = {https://doi.org/10.1103/RevModPhys.89.015007},
  journal={Reviews of Modern Physics},
  volume={89},
  number={1},
  pages={015007},
  year={2017},
  publisher={APS}
}

@article{margueron2018equation,
  title={Equation of state for dense nucleonic matter from metamodeling. I. Foundational aspects},
  author={Margueron, J{\'e}r{\^o}me and Hoffmann Casali, Rudiney and Gulminelli, Francesca},
  doi = {https://doi.org/10.1103/PhysRevC.97.025805},
  journal={Physical Review C},
  volume={97},
  number={2},
  pages={025805},
  year={2018},
  publisher={APS}
}

@article{chen2014building,
  title={Building relativistic mean field models for finite nuclei and neutron stars},
  author={Chen, Wei-Chia and Piekarewicz, J},
  doi = {https://doi.org/10.1103/PhysRevC.90.044305},
  journal={Physical Review C},
  volume={90},
  number={4},
  pages={044305},
  year={2014},
  publisher={APS}
}

@article{huth2021new,
  title={New equations of state constrained by nuclear physics, observations, and QCD calculations of high-density nuclear matter},
  author={Huth, S and Wellenhofer, C and Schwenk, A},
  doi = {https://doi.org/10.1103/PhysRevC.103.025803},
  journal={Physical Review C},
  volume={103},
  number={2},
  pages={025803},
  year={2021},
  publisher={APS}
}

@article{huth2022constraining,
  title={Constraining neutron-star matter with microscopic and macroscopic collisions},
  author={Huth, Sabrina and Pang, Peter TH and Tews, Ingo and Dietrich, Tim and Le F{\`e}vre, Arnaud and Schwenk, Achim and Trautmann, Wolfgang and Agarwal, Kshitij and Bulla, Mattia and Coughlin, Michael W and others},
  doi = {https://doi.org/10.1038/s41586-022-04750-w},
  journal={Nature},
  volume={606},
  number={7913},
  pages={276--280},
  year={2022},
  publisher={Nature Publishing Group UK London}
}

@article{foreman2013emcee,
  title={emcee: the MCMC hammer},
  author={Foreman-Mackey, Daniel and Hogg, David W and Lang, Dustin and Goodman, Jonathan},
  doi = {10.1086/670067},
  journal={Publications of the Astronomical Society of the Pacific},
  volume={125},
  number={925},
  pages={306--312},
  year={2013},
  publisher={University of Chicago Press}
}

@article{lalazissis1997new,
    author = "Lalazissis, G. A. and Konig, J. and Ring, P.",
    title = "{A New parametrization for the Lagrangian density of relativistic mean field theory}",
    eprint = "nucl-th/9607039",
    archivePrefix = "arXiv",
    doi = "10.1103/PhysRevC.55.540",
    journal = "Phys. Rev. C",
    volume = "55",
    pages = "540--543",
    year = "1997"
}

@article{kumar2017new,
  title={New parameterization of the effective field theory motivated relativistic mean field model},
  author={Kumar, Bharat and Singh, SK and Agrawal, BK and Patra, SK},
  doi = {10.1016/j.nuclphysa.2017.07.001},
  journal={Nuclear Physics A},
  volume={966},
  pages={197--207},
  year={2017},
  publisher={Elsevier}
}

@article{kumar2018new,
  title={New relativistic effective interaction for finite nuclei, infinite nuclear matter, and neutron stars},
  author={Kumar, Bharat and Patra, SK and Agrawal, BK},
  doi = {https://doi.org/10.1103/PhysRevC.97.045806},
  journal={Physical Review C},
  volume={97},
  number={4},
  pages={045806},
  year={2018},
  publisher={APS}
}

@article{Das2021impacts,
    author = "Das, H. C. and Kumar, Ankit and Kumar, Bharat and Biswal, S. K. and Patra, S. K.",
    title = "{Impacts of dark matter on the curvature of the neutron star}",
    eprint = "2007.05382",
    archivePrefix = "arXiv",
    primaryClass = "nucl-th",
    doi = "10.1088/1475-7516/2021/01/007",
    journal = "JCAP",
    volume = "01",
    pages = "007",
    year = "2021"
}

@article{stone2014incompressibility,
  title={Incompressibility in finite nuclei and nuclear matter},
  author={Stone, JR and Stone, NJ and Moszkowski, SA},
  doi = {https://doi.org/10.1103/PhysRevC.89.044316},
  journal={Physical Review C},
  volume={89},
  number={4},
  pages={044316},
  year={2014},
  publisher={APS}
}

@article{hornick2018relativistic,
  title={Relativistic parameterizations of neutron matter and implications for neutron stars},
  author={Hornick, Nadine and Tolos, Laura and Zacchi, Andreas and Christian, Jan-Erik and Schaffner-Bielich, J{\"u}rgen},
  doi = {https://doi.org/10.1103/PhysRevC.98.065804},
  journal={Physical Review C},
  volume={98},
  number={6},
  pages={065804},
  year={2018},
  publisher={APS}
}

@article{li2025influence,
  title={Influence of effective nucleon mass on equation of state for supernova simulations and neutron stars},
  author={Li, Shuying and Pang, Junbo and Shen, Hong and Hu, Jinniu and Sumiyoshi, Kohsuke},
  doi = {10.3847/1538-4357/ada6b3},
  journal={The Astrophysical Journal},
  volume={980},
  number={1},
  pages={54},
  year={2025},
  publisher={The American Astronomical Society}
}

@article{you2025u,
  title={U-spin symmetry energy and hyperon puzzle},
  author={You, Hao-Song and Yu, Ting-Lan and Xia, Cheng-Jun and Xu, Ren-Xin},
  eprint = "2511.01325",
  primaryClass = "hep-ph",
  archivePrefix = "arXiv",
  journal={arXiv preprint arXiv:2511.01325},
  year={2025}
}

@article{xia2022unified,
  title={Unified neutron star EOSs and neutron star structures in RMF models},
  author={Xia, Cheng-Jun and Maruyama, Toshiki and Li, Ang and Yuan Sun, Bao and Long, Wen-Hui and Zhang, Ying-Xun},
  doi = {10.1088/1572-9494/ac71fd},
  journal={Communications in Theoretical Physics},
  volume={74},
  number={9},
  pages={095303},
  year={2022},
  publisher={IOP Publishing}
}

@article{bao2015impact,
  title={Impact of the symmetry energy on nuclear pasta phases and crust-core transition in neutron stars},
  author={Bao, SS and Shen, H},
  doi = {https://doi.org/10.1103/PhysRevC.91.015807},
  journal={Physical Review C},
  volume={91},
  number={1},
  pages={015807},
  year={2015},
  publisher={APS}
}

@article{zhu2018neutron,
  title={Neutron star equation of state from the quark level in light of GW170817},
  author={Zhu, Zhen-Yu and Zhou, En-Ping and Li, Ang},
  doi = "10.3847/1538-4357/aacc28",
  journal={The Astrophysical Journal},
  volume={862},
  number={2},
  pages={98},
  year={2018},
  publisher={The American Astronomical Society}
}

@article{170817,
  title={GW170817: Measurements of neutron star radii and equation of state},
  author={Abbott, Benjamin P and Abbott, Richard and Abbott, TD and Acernese, F and Ackley, K and Adams, C and Adams, T and Addesso, P and Adhikari, Rana X and Adya, Vaishali B and others},
  doi = "10.1103/PhysRevLett.121.161101",
  journal={Physical review letters},
  volume={121},
  number={16},
  pages={161101},
  year={2018},
  publisher={APS}
}

@article{danielewicz2002determination,
  title={Determination of the equation of state of dense matter},
  author={Danielewicz, Pawe{\l} and Lacey, Roy and Lynch, William G},
  doi = {10.1126/science.1078070},
  journal={Science},
  volume={298},
  number={5598},
  pages={1592--1596},
  year={2002},
  publisher={American Association for the Advancement of Science}
}

@article{zabari2019influence,
    author = "Zabari, Noemi and Kubis, Sebastian and W{\'o}jcik, W{\l}odzimierz",
    title = "{Influence of the interactions of scalar mesons on the behavior of the symmetry energy}",
    eprint = "1809.03420",
    archivePrefix = "arXiv",
    primaryClass = "nucl-th",
    doi = "10.1103/PhysRevC.99.035209",
    journal = "Phys. Rev. C",
    volume = "99",
    number = "3",
    pages = "035209",
    year = "2019"
}

@article{murakami2001nucleon,
  title={Nucleon scattering with Higgsino and W-ino cold dark matter},
  author={Murakami, Brandon and Wells, James D},
  doi = {https://doi.org/10.1103/PhysRevD.64.015001},
  journal={Physical Review D},
  volume={64},
  number={1},
  pages={015001},
  year={2001},
  publisher={APS}
}

@article{cline2013update,
  title={Update on scalar singlet dark matter},
  author={Cline, James M and Scott, Pat and Kainulainen, Kimmo and Weniger, Christoph},
  doi = {https://doi.org/10.1103/PhysRevD.88.055025},
  journal={Physical Review D—Particles, Fields, Gravitation, and Cosmology},
  volume={88},
  number={5},
  pages={055025},
  year={2013},
  publisher={APS}
}

@article{das2020effects,
  title={Effects of dark matter on the nuclear and neutron star matter},
  author={Das, Harish Chandra and Kumar, Ankit and Kumar, Bharat and Biswal, S Kumar and Nakatsukasa, Takashi and Li, Ang and Patra, SK},
  doi = {https://doi.org/10.1093/mnras/staa1435},
  journal={Monthly Notices of the Royal Astronomical Society},
  volume={495},
  number={4},
  pages={4893--4903},
  year={2020},
  publisher={Oxford University Press}
}

@article{fortin2016neutron,
  title={Neutron star radii and crusts: Uncertainties and unified equations of state},
  author={Fortin, M and Provid{\^e}ncia, C and Raduta, Ad R and Gulminelli, F and Zdunik, JL and Haensel, P and Bejger, M},
  doi = {https://doi.org/10.1103/PhysRevC.94.035804},
  journal={Physical Review C},
  volume={94},
  number={3},
  pages={035804},
  year={2016},
  publisher={APS}
}

@article{baym1971ground,
  title={The ground state of matter at high densities: equation of state and stellar models},
  author={Baym, Gordon and Pethick, Christopher and Sutherland, Peter},
  journal={Astrophysical Journal, vol. 170, p. 299},
  volume={170},
  pages={299},
  year={1971}
}

@article{carriere2003low,
  title={Low-mass neutron stars and the equation of state of dense matter},
  author={Carriere, J and Horowitz, CJ and Piekarewicz, J},
  doi = {10.1086/376515},
  journal={The Astrophysical Journal},
  volume={593},
  number={1},
  pages={463--471},
  year={2003}
}

@article{flanagan2008constraining,
  title={Constraining neutron-star tidal Love numbers with gravitational-wave detectors},
  author={Flanagan, Eanna E and Hinderer, Tanja},
  doi = {https://doi.org/10.1103/PhysRevD.77.021502},
  journal={Physical Review D—Particles, Fields, Gravitation, and Cosmology},
  volume={77},
  number={2},
  pages={021502},
  year={2008},
  publisher={APS}
}

@article{Hinderer:2007mb,
    author = "Hinderer, Tanja",
    title = "{Tidal Love numbers of neutron stars}",
    eprint = "0711.2420",
    archivePrefix = "arXiv",
    primaryClass = "astro-ph",
    doi = "10.1086/533487",
    journal = "Astrophys. J.",
    volume = "677",
    pages = "1216--1220",
    year = "2008",
    note = "[Erratum: Astrophys.J. 697, 964 (2009)]"
}

@article{Damour:2009vw,
    author = "Damour, Thibault and Nagar, Alessandro",
    title = "{Relativistic tidal properties of neutron stars}",
    eprint = "0906.0096",
    archivePrefix = "arXiv",
    primaryClass = "gr-qc",
    doi = "10.1103/PhysRevD.80.084035",
    journal = "Phys. Rev. D",
    volume = "80",
    pages = "084035",
    year = "2009"
}

@article{malik2018gw170817,
  title={GW170817: Constraining the nuclear matter equation of state from the neutron star tidal deformability},
  author={Malik, Tuhin and Alam, N and Fortin, M and Provid{\^e}ncia, C and Agrawal, BK and Jha, TK and Kumar, Bharat and Patra, SK},
  doi = {https://doi.org/10.1103/PhysRevC.98.035804},
  journal={Physical Review C},
  volume={98},
  number={3},
  pages={035804},
  year={2018},
  publisher={APS}
}

@article{Postnikov:2010yn,
    author = "Postnikov, Sergey and Prakash, Madappa and Lattimer, James M.",
    title = "{Tidal Love Numbers of Neutron and Self-Bound Quark Stars}",
    eprint = "1004.5098",
    archivePrefix = "arXiv",
    primaryClass = "astro-ph.SR",
    doi = "10.1103/PhysRevD.82.024016",
    journal = "Phys. Rev. D",
    volume = "82",
    pages = "024016",
    year = "2010"
}

@article{Zhou:2017pha,
    author = "Zhou, En-Ping and Zhou, Xia and Li, Ang",
    title = "{Constraints on interquark interaction parameters with GW170817 in a binary strange star scenario}",
    eprint = "1711.04312",
    archivePrefix = "arXiv",
    primaryClass = "astro-ph.HE",
    doi = "10.1103/PhysRevD.97.083015",
    journal = "Phys. Rev. D",
    volume = "97",
    number = "8",
    pages = "083015",
    year = "2018"
}

@article{Psaltis:2008bb,
    author = "Psaltis, Dimitrios",
    title = "{Probes and Tests of Strong-Field Gravity with Observations in the Electromagnetic Spectrum}",
    eprint = "0806.1531",
    archivePrefix = "arXiv",
    primaryClass = "astro-ph",
    doi = "10.12942/lrr-2008-9",
    journal = "Living Rev. Rel.",
    volume = "11",
    pages = "9",
    year = "2008"
}

@article{Das:2018frc,
    author = "Das, Arpan and Malik, Tuhin and Nayak, Alekha C.",
    title = "{Confronting nuclear equation of state in the presence of dark matter using GW170817 observation in relativistic mean field theory approach}",
    eprint = "1807.10013",
    archivePrefix = "arXiv",
    primaryClass = "hep-ph",
    doi = "10.1103/PhysRevD.99.043016",
    journal = "Phys. Rev. D",
    volume = "99",
    number = "4",
    pages = "043016",
    year = "2019"
}

@article{guichon1988possible,
  title={A possible quark mechanism for the saturation of nuclear matter},
  author={Guichon, Pierre AM},
  doi = {https://doi.org/10.1016/0370-2693(88)90762-9},
  journal={Physics Letters B},
  volume={200},
  number={3},
  pages={235--240},
  year={1988},
  publisher={Elsevier}
}

@article{Saito:2005rv,
    author = "Saito, K. and Tsushima, Kazuo and Thomas, Anthony William",
    title = "{Nucleon and hadron structure changes in the nuclear medium and impact on observables}",
    eprint = "hep-ph/0506314",
    archivePrefix = "arXiv",
    reportNumber = "JLAB-THY-05-326",
    doi = "10.1016/j.ppnp.2005.07.003",
    journal = "Prog. Part. Nucl. Phys.",
    volume = "58",
    pages = "1--167",
    year = "2007"
}

@article{toki1998quark,
  title={Quark mean field model for nucleons in nuclei},
  author={Toki, H and Meyer, U and Faessler, Amand and Brockmann, R},
  journal={Physical Review C},
  doi={https://doi.org/10.1103/PhysRevC.58.3749},
  volume={58},
  number={6},
  pages={3749},
  year={1998},
  publisher={APS}
}

@article{shen2000quark,
  title={Quark mean field model for nuclear matter and finite nuclei},
  author={Shen, H and Toki, H},
  doi={https://doi.org/10.1103/PhysRevC.61.045205},
  journal={Physical Review C},
  volume={61},
  number={4},
  pages={045205},
  year={2000},
  publisher={APS}
}

@article{shen2002study,
  title={Study of $\Lambda$ hypernuclei in the quark mean-field model},
  author={Shen, H and Toki, H},
  doi={10.1016/S0375-9474(02)00961-2},
  journal={Nuclear Physics A},
  volume={707},
  number={3-4},
  pages={469--476},
  year={2002},
  publisher={Elsevier}
}

\end{document}